\newif\ifrevtexavailable
\IfFileExists{revtex4-2.cls}{%
  \revtexavailabletrue
  \documentclass[aps,prb,reprint,superscriptaddress,longbibliography]{revtex4-2}
}{%
  \revtexavailablefalse
  \documentclass[10pt,twocolumn]{article}
  \newcommand{\email}[1]{}
  \newcommand{\affiliation}[1]{}
  \newenvironment{acknowledgments}{\section*{Acknowledgments}}{}
}
\usepackage{upgreek}
\usepackage{amsmath,amssymb,bm,mathtools}
\usepackage{graphicx}
\usepackage{xcolor}
\usepackage{hyperref}

\hypersetup{colorlinks=true,citecolor=blue,linkcolor=blue,urlcolor=blue}

\newcommand{\ii}{\mathrm{i}}
\newcommand{\e}{\mathrm{e}}
\newcommand{\dd}{\mathrm{d}}
\newcommand{\Ms}{M_{\mathrm{s}}}
\newcommand{\Vm}{V_{\mathrm{m}}}
\newcommand{\tf}{t_{\mathrm{F}}}
\newcommand{\ts}{t_{\mathrm{S}}}
\newcommand{\Beff}{\mu_0 M_{\mathrm{eff}}}
\newcommand{\BM}{\mu_0 M_{\mathrm{s}}}
\newcommand{\BH}{\mu_0 H_{\mathrm{ext}}}
\newcommand{\Bex}{\mu_0 H_{\mathrm{ex}}}
\newcommand{\Hone}{\mu_0\mathcal H_1}
\newcommand{\Htwo}{\mu_0\mathcal H_2}
\newcommand{\JS}{\mathbf J_{\mathrm S}}
\newcommand{\AS}{\mathcal A_{\mathrm S}}
\newcommand{\Dif}{\mathcal D}
\newcommand{\Rz}{R_k^{(0)}}
\newcommand{\Rt}{R_{k,s}^{(2)}}
\newcommand{\NlinSC}{N_{\mathrm{SC}}^{\mathrm{lin}}}
\newcommand{\NnlSC}{N_{\mathrm{SC}}^{\mathrm{NL}}}
\newcommand{\Nmag}{N_{\mathrm{mag}}}
\newcommand{\Ntot}{N_{\mathrm{tot}}}

\begin{document}

\title{Chiral magnon Kerr interaction from linear and nonlinear Meissner screening in ferromagnet--superconductor bilayers}

\author{Hanchen Wang}
\email{hanchen.wang@mat.ethz.ch}
\affiliation{Department of Materials, ETH Zurich, 8093 Zurich, Switzerland}

\author{Tao Yu}
\email{taoyuphy@hust.edu.cn}
\affiliation{School of Physics, Huazhong University of Science and Technology, Wuhan 430074, China}

\ifrevtexavailable\else\author{Hanchen Wang and Tao Yu}\fi
\date{\today}

\ifrevtexavailable\else\maketitle\fi

\begin{abstract}
Magnon Kerr interaction showing a dependence of magnon dispersion frequency on its population is a useful functionality for nonlinear and quantum magnonics; its strength and sign depend on magnetic anisotropy, bias, and mode geometry.
We here propose to use a superconducting gate to tune the Kerr interaction of magnons in ferromagnetic insulators.
Even when the superconductor responds linearly, the frequency shift induced by its reflected dipolar field depends on the longitudinal magnetization. The decrease of this magnetization with increasing precession amplitude therefore produces a Kerr contribution.
%In the linear interaction between magnons and a Cooper-pair superfluid, the reflected dipolar field generated by the transverse magnetization and acting on its precession is proportional to the magnetization amplitude, inducing already a  Kerr contribution in the nonlinear regime. 
We go beyond this linear Meissner-screening regime by showing that at finite excitation amplitude, the significant Meissner screening currents reduce the superfluid stiffness, thus enhancing the effective magnetic-field penetration and rendering the Meissner backaction amplitude dependent. This superflow-induced pair breaking adds a second Kerr contribution, controlled by the interfacial superflow relative to the nonlinear screening scale.
For yttrium iron garnet gated by dirty niobium, we predict a significant renormalization of the Kerr effect by superconductors, with the resulting Kerr contribution strongly chiral in the Damon--Eshbach configuration. By showing that the linear and nonlinear Meissner screening to the Kerr interaction can be distinguished by varying temperature, our work establishes a route to directional control of magnon nonlinear interactions.
%In a garnet with tunable perpendicular anisotropy and a purely in-plane, vortex-free bias, counterpropagating spin waves acquire Kerr coefficients of opposite sign.
\end{abstract}

\ifrevtexavailable\maketitle\fi

\section{Introduction}
\label{sec:introduction}

Spin waves in low-damping magnetic insulators carry angular momentum coherently and couple efficiently to microwave circuits~\cite{Serga2010,Chumak2015,Pirro2021}.
In their quantized form, magnons are building blocks of hybrid quantum systems, in which controllable magnon interactions enable nonlinear and quantum functionalities~\cite{Li2020Hybrid,Yuan2022,Rameshti2022,Elyasi2020}.
The leading self-interaction of a magnon mode is the Kerr interaction, which shifts the mode frequency in proportion to its magnon population.
It governs nonlinear relaxation, phase locking, and frequency combs~\cite{Mohseni2021,Dreyer2022,Wang2024Comb}, and underlies nonreciprocity, multistability, squeezing, entanglement, and magnon blockade~\cite{Kong2019,Wang2018PRL,Bi2021,Yang2021,Zou2020,Hou2024}.
Its origin is the nonlinearity of magnetization dynamics~\cite{AndersonSuhl1955,Suhl1957,Lvov1994,Krivosik2010}.
In cavity magnonics, this nonlinearity is attributed to magnetic anisotropy~\cite{Wang2016PRB,Zhang2019Theory}, and recent thin-film experiments show that its magnitude and sign depend on magnetization direction and precession ellipticity~\cite{Petrosyan2026Exp,Petrosyan2026Theory}, consistent with the angular dependence of the nonlinear frequency shift in spin-torque oscillators~\cite{SlavinTiberkevich2009}.
In these cases, the Kerr coefficient depends on the magnetic anisotropy, bias, and mode geometry.
A Kerr coefficient that can be engineered through the electromagnetic environment and that differs for counterpropagating waves would open new functionalities, analogous to Kerr-based nonreciprocal devices in photonics~\cite{DelBino2018}.

A superconductor (SC) adjacent to a ferromagnetic insulator (FI) provides such an environment~\cite{yu2026electromagnetic}.
The dipolar electromagnetic field emitted by a spin wave drives Meissner screening currents, which reflect the electromagnetic field that acts back on the magnet.
In the linear regime, this backaction renormalizes the spin-wave dispersion and makes it nonreciprocal~\cite{Golovchanskiy2018,Kuznetsov2022,yu2022efficient,Zhou2024PRBL}, and hybrid spin-wave--Meissner-current modes have been imaged directly~\cite{Borst2023}.
Related effects include superconducting gating of ferromagnetic resonance~\cite{zhou2023gating}, superconductivity-dependent magnon--photon hybridization~\cite{Qiu2024PRB,Ghirri2024}, Meissner-mediated coupling between magnets~\cite{MeissnerCoupling2025}, and superconducting magnonic landscapes~\cite{Golovchanskiy2019,Kharlan2024,Fluxonics2022,Dobrovolskiy2019NP,Dobrovolskiy2025NN}.
FI--SC hybrid circuits reach the strong-coupling and quantum regimes at millikelvin temperatures~\cite{Huebl2013,Tabuchi2015,YIGSuperconducting2017,Li2019PRL,Li2022PRL,YIGCircuit2025,SingleShot2025}.
The SC is, moreover, a nonlinear medium.
A supercurrent reduces the superfluid density and thus the screening it produces.
This nonlinear Meissner effect, first discussed for nodal superconductors~\cite{Yip1992PRL,Xu1995PRB}, appears in conventional films as a current-dependent surface reactance and kinetic inductance~\cite{Gittleman1965,Dahm1997JAP,Groll2010NLME,Nb3Sn2022,DirtyFlow2020,Sauls2022PTEP}.
It provides the Kerr terms of superconducting resonators and kinetic-inductance amplifiers~\cite{Zmuidzinas2012,Eom2012,Maleeva2018,Shu2021NbTiN}.
Nonlinear spin waves in FI--SC hybrids, however, have been studied with a linearly responding SC~\cite{Golovchanskiy2020}, and how the Meissner backaction affects the magnon Kerr interaction has remained open.

In this work, we derive the Kerr coefficient of spin waves in an FI$|$SC bilayer, particularly focusing on the Damon--Eshbach  configuration~\cite{Damon1960,Damon1961,Kalinikos1986}, proceeding from the bare FI to linear and then nonlinear Meissner screening regimes.
We find that the SC contributes through two channels of comparable size.
In the linear Meissner-screening channel, the reflected electromagnetic field is linear in the transverse magnetization, but its effect on the precession scales with the longitudinal magnetization, which decreases with increasing precession amplitude.
In the nonlinear Meissner-screening channel, a superflow-induced pair breaking weakens the screening itself.
Both channels are chiral, because counterpropagating spin waves in the Damon--Eshbach configuration load the two FI surfaces differently~\cite{Yu2023Chirality,Chirality2024,Unidirectional2022,BilayerNonreciprocity2025}, but they differ by orders of magnitude in their directional contrast and in their temperature dependence, providing complementary signatures for distinguishing them experimentally.
Finally, we show that in a garnet with tunable perpendicular anisotropy~\cite{wang2025ultrathin,Soumah2018}, the Kerr coefficients of counterpropagating waves can be made opposite in sign under an in-plane bias compatible with the Meissner state.
For a representative low-temperature heterostructure of Nb and strained, Bi-substituted yttrium iron garnet (YIG), we consider the anisotropy-compensation point $\mu_0M_{\mathrm{eff}}=0$~\cite{wang2025ultrathin,Soumah2018}. At $|k|=1.81\,\mathrm{rad}/\upmu\mathrm m$ and volume $\Vm=25\,\upmu\mathrm m^3$, the bare-film Kerr coefficient is $\mathcal K_{\mathrm{bare}}/(2\pi)\simeq+0.62\,\mathrm{mHz}$. Superconducting screening changes it to $\mathcal K/(2\pi)\simeq-0.87$ and $+0.59\,\mathrm{mHz}$ for the $+k$ and $-k$ spin-wave modes, respectively, demonstrating an efficient and chiral control of the Kerr effect by SCs.

The paper is organized as follows.
Section~\ref{sec:model} introduces the model.
Section~\ref{sec:linear} treats linear spin waves, first in the bare FI and then with Meissner screening.
Section~\ref{sec:kerr_bare} derives the Kerr coefficient of the bare film, and Sec.~\ref{sec:kerr_linear} adds the linear-screening contribution.
Section~\ref{sec:nonlinear_meissner} develops the nonlinear Meissner response and its contribution.
Section~\ref{sec:compensation} discusses directional Kerr compensation, Sec.~\ref{sec:quantum_correspondence} maps the results onto the quantum Kerr interaction, and Sec.~\ref{sec:conclusion} discusses experimental tests and the range of validity and concludes.

\section{Model}
\label{sec:model}

\begin{figure}[htp!]
\centering
\includegraphics[width=1.02\columnwidth]{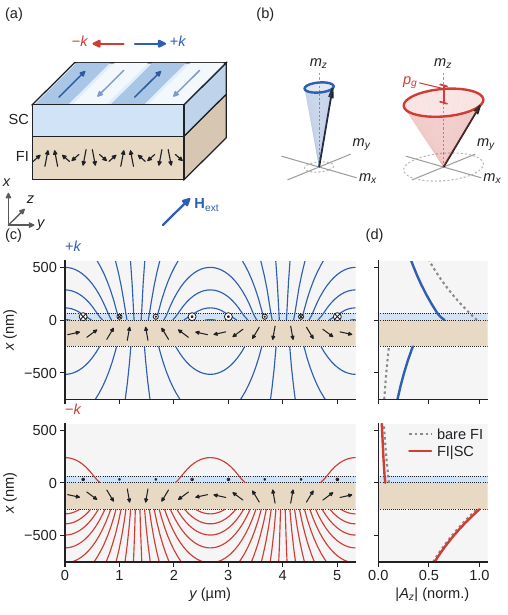}
\caption{(a) FI$|$SC bilayer.
The bias field $\mathbf H_{\rm ext}$ and the equilibrium magnetization lie in the film plane along $\hat{\mathbf z}$, and spin waves in the Damon--Eshbach configuration propagate along $\pm\hat{\mathbf y}$.
Arrows in the FI indicate the dynamic magnetization; stripes on the SC surface indicate the screening current $J_{{\rm S},z}$.
(b) Precession at vanishing and at finite amplitude.
The bar marks the fractional reduction of the longitudinal magnetization, $p_g=1-\langle M_m\rangle/\Ms$, which measures the spin-wave intensity.
(c) Distribution of magnetic stray fields, i.e., contours of $A_z(x,y)$, of the $+k$ (upper) and $-k$ (lower) modes over one wavelength, at equal intensity and with the same contour spacing.
The FI interior shows the dynamic magnetization instead of field lines.
Symbols in the SC mark the screening current along $+\hat{\mathbf z}$ ($\otimes$) and $-\hat{\mathbf z}$ ($\odot$), with an area proportional to its magnitude.
(d) Envelope $|A_z(x)|$ for the FI$|$SC bilayer (solid) and for the bare FI (dotted), normalized to a common maximum; the FI interior is omitted.
Background colors: vacuum (gray), FI (tan), SC (blue).
Panels (c,d) use $|k|=1.18\,\mathrm{rad}/\upmu\mathrm m$, $T=0.1\,T_c$ with $T_c$ the superconducting critical temperature, $\tf=250\,\mathrm{nm}$, $\ts=65\,\mathrm{nm}$, and the YIG$|$Nb parameters of Secs.~\ref{sec:linear} and \ref{sec:nonlinear_meissner} (Appendix~\ref{app:fig1_profile}).}
\label{fig:principle}
\end{figure}

Figure~\ref{fig:principle}(a) shows an FI of thickness $\tf$, occupying $-\tf<x<0$, covered by an SC of thickness $\ts$ in $0<x<\ts$, where $x$ is the coordinate along the film normal.
An external field $\mathbf H_{\rm ext}=H_{\rm ext}\hat{\mathbf z}$ is applied in the film plane.
We consider an FI mode propagating along $y$, perpendicular to the field (i.e., the Damon--Eshbach geometry), with signed wave vector $k=s|k|$, where $s=\operatorname{sgn}k=\pm1$ labels the propagation direction.

The magnetic anisotropy is characterized by the effective magnetization
\begin{equation}
 \Beff=\BM-\frac{2K_\perp}{\Ms},
 \label{eq:Meff}
\end{equation}
where $\mu_0$ is the vacuum permeability, $\Ms$ is the saturation magnetization, the first term is the thin-film shape anisotropy, and $K_\perp$ is a uniaxial magnetic anisotropy constant favoring the film normal. 
In Bi-substituted YIG, composition, growth conditions, and epitaxial strain provide control over the perpendicular anisotropy, allowing compensation and reversal of the effective anisotropy~\cite{wang2025ultrathin,Soumah2018}. In our uniaxial model, the compensation point $\Beff=0$ corresponds to $K_\perp=\mu_0\Ms^2/2$.
Anisotropy compensation at $M_{\mathrm{eff}}=0$ yields circular uniform-mode precession in the bare film~\cite{Divinskiy2019Nonlinear}, suppressing ellipticity-induced nonlinear magnon scattering and facilitating antimagnon excitation when an inverted magnetization state is stabilized by spin-orbit torque~\cite{Wang2026Antimagnons,Karadza2026Dynamical,Kurebayashi2026Dynamical}.
For $\BH+\Beff>0$, the equilibrium magnetization lies along $\hat{\mathbf e}_m=\hat{\mathbf z}$.
A uniformly magnetized film of infinite lateral extent produces no static stray field outside, irrespective of $\Beff$, so the SC experiences only the in-plane field $B_\parallel=\BH$.

With the right-handed transverse fluctuations along $\hat{\mathbf e}_1=\hat{\mathbf y}$ and $\hat{\mathbf e}_2=-\hat{\mathbf x}$, we express the magnetization 
\begin{equation}
\mathbf M=M_m\hat{\mathbf e}_m+\sum_{j=1,2}\operatorname{Re}\big[\mathcal M_j\e^{\ii(ky-\omega t)}\big]\hat{\mathbf e}_j,
 \label{eq:M_decomposition}
\end{equation}
where $\mathcal M_j$ are complex transverse magnetization amplitudes with units of $\mathrm{A/m}$.
The real transverse components are $\delta M_j=\operatorname{Re}[\mathcal M_j\e^{\ii(ky-\omega t)}]$, and the longitudinal magnetization $M_m=\sqrt{\Ms^2-\delta M_1^2-\delta M_2^2}$ enforces the fixed magnetization length.
The magnetic energy
\begin{equation}
 \begin{aligned}
 \mathcal E={}&-\mu_0\int_{\Vm}\mathbf M\cdot\mathbf H_{\rm ext}\,\dd V
 +\frac{\Beff}{2\Ms}\int_{\Vm}M_x^2\,\dd V\\
 &+\frac{A_{\rm ex}}{\Ms^2}\int_{\Vm}\sum_{i=x,y,z}(\nabla M_i)^2\dd V
 -\frac{\mu_0}{2}\int_{\Vm}\mathbf M\cdot\mathbf H_d\,\dd V
 \end{aligned}
 \label{eq:classical_energy_full}
\end{equation}
contains, in this order, the Zeeman energy, the effective uniaxial anisotropy energy (including the static demagnetizing energy through $\Beff$), the exchange energy with exchange stiffness $A_{\rm ex}$, and the energy of the residual dipolar field $\mathbf H_d$ generated by the spin wave.
The local shape-demagnetizing contribution is already included in $\Beff$ and is excluded from $\mathbf H_d$ to avoid double counting.
This quadratic energy describes the bare FI and linear screening; the nonlinear SC backaction is introduced through its current--field relation in Sec.~\ref{sec:nonlinear_meissner}.
Here, $\Vm$ is the magnetic volume, $M_x$ is the magnetization component along the film normal, and $M_i$ are the Cartesian components of $\mathbf M$.
The lossless magnetization dynamics follow the Landau--Lifshitz equation
\begin{equation}
 \frac{\partial\mathbf M}{\partial t}=-\gamma\mu_0\mathbf M\times\mathbf H_{\rm eff},
 \qquad
 \mu_0\mathbf H_{\rm eff}=-\frac{\delta\mathcal E}{\delta\mathbf M},
 \label{eq:LL_classical}
\end{equation}
where $\gamma>0$ is the magnitude of the gyromagnetic ratio and $\mathbf H_{\rm eff}$ the effective magnetic field.
Damping does not affect the conservative frequency shift and is neglected.

As a measure of the spin-wave intensity, we use the fractional reduction of the longitudinal magnetization [Fig.~\ref{fig:principle}(b)],
\begin{equation}
p_g\equiv1-\frac{\langle M_m\rangle}{\Ms}
=\frac{|\mathcal M_1|^2+|\mathcal M_2|^2}{4\Ms^2}+O(p_g^2),
 \label{eq:pg}
\end{equation}
where $\langle\cdots\rangle$ denotes the average over one period and the subscript ``$g$" indicates a global, mode-averaged quantity.
For circular precession with cone angle $\theta_{\rm c}$, $p_g=1-\cos\theta_{\rm c}$.

\section{Linear spin waves in FI-SC heterostructure}
\label{sec:linear}

\subsection{Bare ferromagnetic film}
\label{sec:linear_bare}

We approximate the FI mode as uniform across its thickness and average the dipolar field accordingly~\cite{Kalinikos1986}. This approximation neglects coupling to higher thickness modes.
In the complex-amplitude representation, the thickness-averaged dipolar field is $\overline{\mathbf h}_{xy} =\mathbf D_0\,\boldsymbol{\mathcal M}_{xy}$, where $\overline{\mathbf h}_{xy}$ denotes its complex amplitude and $\boldsymbol{\mathcal M}_{xy}=(\mathcal M_x,\mathcal M_y)^T$, with $\mathcal M_x=-\mathcal M_2$ and $\mathcal M_y=\mathcal M_1$.
The dipolar tensor is
\begin{equation}
 \mathbf D_0=\begin{pmatrix}-F_k&0\\0&F_k-1\end{pmatrix},\qquad
 F_k=\frac{1-\e^{-|k|\tf}}{|k|\tf},
 \label{eq:D0}
\end{equation}
where $F_k$ is the demagnetizing factor of a normally polarized magnetization modulation with in-plane wavelength $2\pi/|k|$, which decreases from $1$ at $k=0$ as the magnetic charges of neighboring half-wavelengths compensate.
Because $\Beff$ already contains the static ($k=0$) demagnetizing field, only the finite-$k$ part of $\mathbf D_0$ enters explicitly.
The transverse field acting on $\mathcal M_j$ is then $b_j^{\rm m}\mathcal M_j/\Ms$, with the self-field coefficients
\begin{equation}
 \begin{aligned}
 b_1^{\rm m}&=-\BM(1-F_k)-\Bex,\\
 b_2^{\rm m}&=-\Beff+\BM(1-F_k)-\Bex,
 \end{aligned}
 \label{eq:selffields_bare}
\end{equation}
where $\Bex=2A_{\rm ex}k^2/\Ms$ is the exchange field.
Linearizing Eq.~(\ref{eq:LL_classical}) gives the reciprocal spin-wave dispersion in the Damon--Eshbach configuration 
\begin{equation}
 \omega_k^{(0)}=\gamma\sqrt{\Hone\Htwo},\qquad
 \mu_0\mathcal H_j=\BH-b_j^{\rm m},
 \label{eq:omega_bare}
\end{equation}
where the stiffness fields $\mu_0\mathcal H_j$ are the curvatures of the magnetic energy density divided by $\Ms$ for tilts toward $\hat{\mathbf e}_j$, and the superscript $(0)$ denotes the linear (vanishing intensity) limit.
Because $\mathcal H_1\neq\mathcal H_2$ in general, the precession is elliptical, with ellipticity and normal-mode factor
\begin{equation}
 a_k=\left|\frac{\mathcal M_2}{\mathcal M_1}\right|=\sqrt{\frac{\Hone}{\Htwo}},\qquad
 \nu_k=\frac{1}{2}\left(a_k+\frac{1}{a_k}\right).
 \label{eq:ellipticity}
\end{equation}
The factor $\nu_k=|u_k|^2+|v_k|^2\geq1$, where $u_k$ and $v_k$ are the coefficients of the Bogoliubov transformation that diagonalizes the quadratic magnon Hamiltonian, is the number of spin flips carried by one magnon of the elliptical mode~\cite{Zou2020,Yuan2022}.
To leading order in $p_g$, Eq.~(\ref{eq:pg}) gives $|\mathcal M_1|=2\Ms[p_g/(1+a_k^2)]^{1/2}$ and $|\mathcal M_2|=a_k|\mathcal M_1|$.

For numerical estimates we consider a YIG film with $\tf=250\,\mathrm{nm}$ in $\BH=0.30\,\mathrm T$, with the low-temperature saturation magnetization $\BM=0.245\,\mathrm T$ ~\cite{GurevichMelkov1996}, $A_{\rm ex}=3.6\,\mathrm{pJ/m}$~\cite{wang2020chiral}, and $\gamma/(2\pi)=28\,\mathrm{GHz/T}$.
Unless stated otherwise, only shape anisotropy is included since $K_{\perp}/M_s\ll \mu_0 M_s$, i.e., $\Beff=\BM$. 

\subsection{Renormalization by Meissner screening of SC}
\label{sec:linear_SC}

The dipolar stray field of the spin wave decays outside the FI as $\exp(-|k||x|)$.
At the upper interface, it drives Meissner currents in the SC, which produce a reflected evanescent field acting back on the FI.
In general, this backaction follows from matching the electric and magnetic fields at all interfaces~\cite{zhou2023gating,Zhou2024PRBL,yu2026electromagnetic}.
In the quasistatic limit, it reduces to a wave-vector-dependent reflection coefficient.

\begin{figure}[!t]
\centering
\includegraphics[width=\columnwidth]{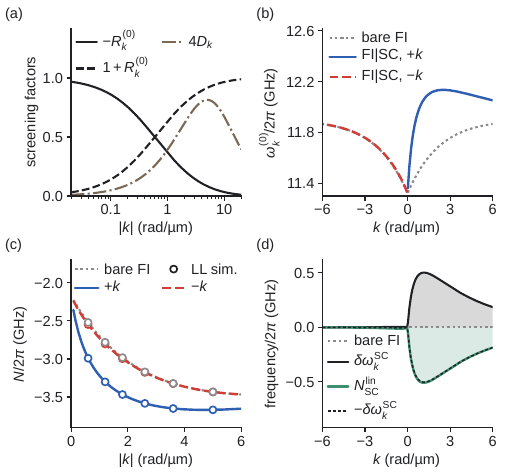}
\caption{Linear Meissner screening for the YIG$|$Nb bilayer with $\tf=250\,\mathrm{nm}$, $\ts=65\,\mathrm{nm}$, $\BH=0.30\,\mathrm T$ in plane, $\Beff=\BM$, and $T=0.1\,T_c$.
(a)~Factors of the reflected field $G_k=\BM D_k\Rz$: screening strength $-\Rz$, field penetrating the SC, $1+\Rz$, and overlap $4D_k$.
The product $D_k\Rz$ vanishes for $|k|\tf\rightarrow0$ and for $|k|\lambda\gg1$.
The bare FI has $\Rz=0$.
(b)~Linear dispersion of the FI$|$SC bilayer for $+k$ (solid) and $-k$ (dashed), and of the bare FI (dotted).
(c)~$\Nmag^{(k)}+\NlinSC$ for $\pm k$ and the intrinsic coefficient $\Nmag^{(k)}$ of the bare FI, Eq.~(\ref{eq:Nmag_k}), which approaches the macrospin value $\Nmag^{(0)}/2\pi=-2.18\,\mathrm{GHz}$ for $k\rightarrow0$.
Symbols: direct integration of the Landau--Lifshitz equation (Appendix~\ref{app:harmonic_balance}).
(d)~Linear Meissner frequency shift $\delta\omega_k^{\rm SC}$ [difference between each bilayer branch and the bare-film curve in (b)] and Kerr contribution of linear screening $\NlinSC$, Eq.~(\ref{eq:Nlin}).
For $+k$, the curve $-\delta\omega_k^{\rm SC}$ (dotted black) nearly coincides with $\NlinSC$, consistent with Eq.~(\ref{eq:Nlin_sumrule}); the relative deviation for $-k$ is discussed in Sec.~\ref{sec:kerr_linear}.}
\label{fig:linear}
\end{figure}

For the dipolar fields in the Damon--Eshbach configuration, the vector potential has only a $z$ component, $\mathbf A=\mathcal A(x)\hat{\mathbf z}\exp[\ii(ky-\omega t)]$.
Inside the SC, the London equation gives
\begin{equation}
 \big(\partial_x^2-[q_k^{(0)}]^2\big)\mathcal A=0,\qquad
 q_k^{(0)}=\sqrt{k^2+\lambda^{-2}},
 \label{eq:qlinear}
\end{equation}
where $\lambda\equiv\lambda(T)$ is the zero-current magnetic penetration depth at temperature $T$, so that $1/q_k^{(0)}$ is the penetration length of the Fourier component $k$.
Outside the SC, the field decays as $\exp[-|k|(x-\ts)]$, which imposes $\partial_x\mathcal A(\ts)=-|k|\mathcal A(\ts)$.
The response at the FI--SC interface is then characterized by the logarithmic derivative
\begin{equation}
 \Lambda_k^{(0)}=-\left.\frac{\partial_x\mathcal A}{\mathcal A}\right|_{x=0^+}
 =q_k^{(0)}\frac{|k|+q_k^{(0)}\tanh(q_k^{(0)}\ts)}{q_k^{(0)}+|k|\tanh(q_k^{(0)}\ts)}.
 \label{eq:Lambdafinite}
\end{equation}
$\Lambda_k^{(0)}$ is the wave-vector-resolved magnetic surface admittance of the SC.
With the electric field $E_z=\ii\omega A_z$, where $A_z$ is the $z$ component of $\mathbf A$, and the tangential magnetic field $H_y=\Lambda_k^{(0)}A_z/\mu_0$ at the surface, the quasistatic surface impedance is $Z_{\rm S}=E_z/H_y=\ii\omega\mu_0/\Lambda_k^{(0)}$, which reduces to the familiar surface reactance $X_{\rm S}=\omega\mu_0\lambda$ for $k\rightarrow0$ and $\ts\gg\lambda$~\cite{Tinkham1996,Hein1999}.
Matching the incident and reflected evanescent fields to this boundary condition yields the reflection coefficient (Appendix~\ref{app:linear_reflection})
\begin{equation}
 \Rz=\frac{|k|-\Lambda_k^{(0)}}{|k|+\Lambda_k^{(0)}},
 \label{eq:Rlinear}
\end{equation}
defined as the ratio of the reflected to the incident amplitude of the vector potential.
It approaches $-1$ (perfect screening) for $|k|\rightarrow0$ and vanishes for $|k|\lambda\gg1$ or $\ts\rightarrow0$ [see Fig.~\ref{fig:linear}(a) below].

The reflected field adds to the thickness-averaged dipolar field the reflected tensor
\begin{equation}
 \mathbf D_{\rm R}=D_k\Rz\begin{pmatrix}1&-\ii s\\ \ii s&1\end{pmatrix},\qquad
 D_k=\frac{F_k}{2}\big(1-\e^{-|k|\tf}\big),
 \label{eq:DR}
\end{equation}
where $D_k$ is the overlap between the dipolar field of the FI and that of its Meissner reflected field.
The off-diagonal elements of $\mathbf D_{\rm R}$ are odd in $k$: because the transverse components of a positive-frequency spin wave are in quadrature, their stray fields add at the upper FI surface for one propagation direction and cancel for the other [Fig.~\ref{fig:principle}(c,d)].
The reflected field shifts both self-field coefficients,
\begin{equation}
 b_j=b_j^{\rm m}+G_k,\qquad G_k=\BM D_k\Rz\leq0,
 \label{eq:selffields}
\end{equation}
where $G_k$ is the reflected field per unit normalized magnetization.
In addition, $G_k$ couples $\mathcal M_1$ and $\mathcal M_2$ through the chiral off-diagonal term.
The dispersion becomes
\begin{equation}
 \omega_k^{(0)}(s)=\gamma\Big[\sqrt{\Hone\Htwo}-s\,G_k\Big],\quad
 \mu_0\mathcal H_j=\BH-b_j,
 \label{eq:omega_linear}
\end{equation}
which has the Damon--Eshbach form of the linear FI$|$SC problem~\cite{Zhou2024PRBL}, while $a_k$ and $\nu_k$ retain the form of Eq.~(\ref{eq:ellipticity}) with the screened stiffness fields.
The odd term $-sG_k$ stiffens the $+k$ branch [Fig.~\ref{fig:linear}(b)].
To first order in $G_k$, the SC shifts the frequency by
\begin{equation}
 \delta\omega_k^{\rm SC}(s)\simeq-\gamma G_k(\nu_k+s),\qquad
 \nu_k+s=\frac{(1+s\,a_k)^2}{2a_k},
 \label{eq:linear_SC_shift}
\end{equation}
which we call the linear Meissner frequency shift.
The identity in Eq.~(\ref{eq:linear_SC_shift}) shows that the shift is proportional to $(1+s\,a_k)^2$, i.e., to the squared amplitude of the stray field at the SC side.

%For the SC we take $\ts=65\,\mathrm{nm}$ and $\lambda(0)=226\,\mathrm{nm}$, representative of a disordered Nb film (Sec.~\ref{sec:nonlinear_London}), and $T=0.1\,T_c$, where $T_c$ is the superconducting critical temperature and $\lambda\simeq\lambda(0)$. The counterpropagating splitting $\omega_k^{(0)}(+)-\omega_k^{(0)}(-)$ then reaches $2\pi\times0.50\,\mathrm{GHz}$ at $|k|=1.17\,\mathrm{rad}/\upmu\mathrm m$ [Fig.~\ref{fig:linear}(a)].

\section{Kerr coefficient of bare film}
\label{sec:kerr_bare}

At finite intensity, the mode frequency depends on $p_g$,
\begin{equation}
 \omega_k(p_g,s)=\omega_k^{(0)}(s)+N(k,s)\,p_g+O(p_g^2),
 \label{eq:classical_frequency_expansion}
\end{equation}
which defines the \emph{nonlinear frequency-shift coefficient} 
\[
N\equiv\partial\omega_k/\partial p_g|_{p_g=0}.
\]
Since $p_g$ is dimensionless, $N$ has units of angular frequency. $N$ characterizes a nonlinear frequency shift that depends on the wave amplitudes. Its relation to the magnon Kerr coefficient becomes clear in Sec.~\ref{sec:quantum_correspondence} below.
The nonlinear shift arises from the product of each transverse self-field with the longitudinal magnetization $M_m$ in Eq.~(\ref{eq:LL_classical}).
Its period average, $\langle M_m\rangle=\Ms(1-p_g)$, decreases with intensity.
Elliptical precession also produces an oscillating longitudinal component, which we retain in the harmonic-balance calculation below.

We evaluate $N$ by harmonic balance.
We use the normalized magnetization $\mathbf m=\mathbf M/\Ms$ and the phase $\vartheta=ky-\omega t$.
Choosing the overall phase appropriately, the transverse fundamental is $m_x=\bar m_x\cos\vartheta$ and $m_y=-\bar m_y\sin\vartheta$, with dimensionless real amplitudes $\bar m_x=|\mathcal M_2|/\Ms$ and $\bar m_y=|\mathcal M_1|/\Ms$.
Thus, $\bar m_x/\bar m_y=a_k$ and $\bar m_x^2+\bar m_y^2=4p_g$ to leading order.
The longitudinal component is
\begin{equation}
 m_z=1-\frac{1}{2}\big(\Sigma_0+\Xi\cos2\vartheta\big)+O(p_g^2),
 \label{eq:mz_expansion}
\end{equation}
with the static part $\Sigma_0=(\bar m_x^2+\bar m_y^2)/2$ and the second-harmonic amplitude $\Xi=(\bar m_x^2-\bar m_y^2)/2$, which is nonzero for elliptical precession.
Projecting the products of $m_z$ with the transverse fields onto the fundamental harmonic reduces a field component in phase with $m_x$ by the factor $c_x=1-(3\bar m_x^2+\bar m_y^2)/8$, and one in phase with $m_y$ by $c_y=1-(\bar m_x^2+3\bar m_y^2)/8$.
In addition, the second harmonic of $m_z$ generates the exchange field $2\Bex\,\Xi\cos2\vartheta\,\hat{\mathbf z}$; it generates no dipolar field because the magnetization does not vary along $z$.
Inserting these results into Eq.~(\ref{eq:LL_classical}) gives
\begin{equation}
 \begin{aligned}
 \omega\,\bar m_x&=\gamma\big(\BH-\Bex\Xi-c_yb_1^{\rm m}\big)\bar m_y,\\
 \omega\,\bar m_y&=\gamma\big(\BH+\Bex\Xi-c_xb_2^{\rm m}\big)\bar m_x,
 \end{aligned}
 \label{eq:HB_bare}
\end{equation}
which reduce to Eq.~(\ref{eq:omega_bare}) for $c_x=c_y=1$ and $\Xi=0$.
Solving Eq.~(\ref{eq:HB_bare}) to first order in $p_g$ (Appendix~\ref{app:harmonic_balance}) yields the intrinsic coefficient of the finite-$k$ mode,
\begin{equation}
\begin{aligned}
 \frac{\Nmag^{(k)}}{\gamma}
 &=\frac{(a_k^2+3)\,b_1^{\rm m}+a_k^2(3a_k^2+1)\,b_2^{\rm m}}{4a_k(1+a_k^2)}\\
 &\quad+\Bex\frac{(1-a_k^2)^2}{a_k(1+a_k^2)}.
 \end{aligned}
  \label{eq:Nmag_k}
 \end{equation}
The first term collects the weakening of the anisotropy, finite-$k$ dipolar, and exchange restoring fields, weighted by the ellipticity; the second arises from the exchange field of the longitudinal second harmonic.
Within the single-thickness-mode model, Eq.~(\ref{eq:Nmag_k}) is exact to first order in $p_g$, which we confirm by direct numerical integration of Eq.~(\ref{eq:LL_classical}) (Appendix~\ref{app:harmonic_balance}).

In the macrospin limit $k\rightarrow0$, $b_1^{\rm m}=0$, $b_2^{\rm m}=-\Beff$, and $a_k^2=\BH/(\BH+\Beff)$, so that Eq.~(\ref{eq:Nmag_k}) reduces to the classical in-plane result~\cite{SlavinTiberkevich2009}
\begin{equation}
 \Nmag^{(0)}=-\frac{\gamma\Beff}{2\nu_0}
 \left[1-\frac{3\Beff}{4(\BH+\Beff)}\right],
 \label{eq:NIP_classical}
\end{equation}
where $\nu_0$ is the normal-mode factor of the uniform mode.
The mode thus softens with intensity in an easy-plane film ($\Beff>0$) and hardens when the perpendicular anisotropy dominates ($\Beff<0$).
The sign change at $\Beff\approx0$ is the basis of the compensation scheme of Sec.~\ref{sec:compensation}.
At the wave vectors of interest, however, the macrospin value is inaccurate: for $\Beff=\BM$ and $|k|=1.81\,\mathrm{rad}/\upmu\mathrm m$, Eq.~(\ref{eq:Nmag_k}) gives $\Nmag^{(k)}/2\pi=-2.98\,\mathrm{GHz}$, compared with $-2.18\,\mathrm{GHz}$ from Eq.~(\ref{eq:NIP_classical}) [Fig.~\ref{fig:linear}(c)].
The difference stems from the finite-$k$ dipolar fields and exceeds the superconducting contributions derived below, so the finite-$k$ intrinsic coefficient should be used for a consistent comparison.

\section{Kerr contribution from linear Meissner screening}
\label{sec:kerr_linear}

Here, ``linear screening'' refers to the SC response to the transverse magnetization. The fixed-length constraint on the magnetization still makes its dynamics nonlinear. The reflected field enters the equations of motion in the same way as the magnetic self-fields.
According to Eq.~(\ref{eq:selffields}), it adds $G_k$ to both self-field coefficients and introduces the chiral coupling, so that Eq.~(\ref{eq:HB_bare}) becomes
\begin{equation}
 \begin{aligned}
 \big(\omega+\gamma c_ysG_k\big)\bar m_x&=\gamma\big(\BH-\Bex\Xi-c_yb_1\big)\bar m_y,\\
 \big(\omega+\gamma c_xsG_k\big)\bar m_y&=\gamma\big(\BH+\Bex\Xi-c_xb_2\big)\bar m_x.
 \end{aligned}
 \label{eq:HB_SC}
\end{equation}
The reflected field is generated by the transverse magnetization, but, like all self-fields, it acts on the precession in proportion to $M_m$, which is why it carries the reduction factors $c_{x,y}$ even though the SC responds linearly.
To first order in $p_g$, Eq.~(\ref{eq:HB_SC}) yields $N=\Nmag^{(k)}+\NlinSC$, where $\Nmag^{(k)}$ is given by Eq.~(\ref{eq:Nmag_k}) evaluated with the screened ellipticity and
\begin{equation}
 \NlinSC(k,s)=\gamma G_k\big(\nu_k+s+\chi_k\big),\qquad
 \chi_k=\frac{(1-a_k^2)^2}{4a_k(1+a_k^2)}.
 \label{eq:Nlin}
\end{equation}
The correction $\chi_k\geq0$ stems from the second harmonic of $M_m$ and vanishes for circular precession.
For the $+k$ branch, neglecting $\chi_k$ relative to $\nu_k+1$ and keeping the leading order in $G_k$ gives
\begin{equation}
 \NlinSC(k,+1)\simeq-\delta\omega_k^{\rm SC}(+1).
 \label{eq:Nlin_sumrule}
\end{equation}
This relation expresses the weakening of Meissner stiffening as the longitudinal magnetization decreases.
It is not uniformly accurate for the $-k$ branch: near circular precession, $\chi_k\simeq\nu_k-1$, so the correction is comparable to the $-k$ coupling itself.
At $|k|=1.81\,\mathrm{rad}/\upmu\mathrm m$, $-\NlinSC/\delta\omega_k^{\rm SC}=1.007$ for $+k$ but $1.953$ for $-k$ [Fig.~\ref{fig:linear}(d)].
As a consequence, $\NlinSC\leq0$, and its directional contrast follows from Eqs.~(\ref{eq:linear_SC_shift}) and (\ref{eq:Nlin}),
\begin{equation}
 \frac{\NlinSC(-|k|)}{\NlinSC(+|k|)}=\left(\frac{1-a_k}{1+a_k}\right)^2
 \frac{3+2a_k+3a_k^2}{3-2a_k+3a_k^2},
 \label{eq:Nlin_ratio}
\end{equation}
which inherits the factor $(1\pm a_k)^2$ of the interfacial stray field.

For the YIG$|$Nb parameters, the contribution to the Kerr interaction from the $+k$ spin waves by the linear Meissner screening peaks at $\NlinSC/2\pi=-0.51\,\mathrm{GHz}$ near $|k|=1.1\,\mathrm{rad}/\upmu\mathrm m$ [Fig.~\ref{fig:kerr}(b)] and amounts to $16\%$ of $|\Nmag^{(k)}|$ at $|k|=1.81\,\mathrm{rad}/\upmu\mathrm m$.
It peaks at an intermediate $|k|$ because the reflected field, $\propto D_k\Rz$ [Fig.~\ref{fig:linear}(a)], vanishes both for $|k|\tf\rightarrow0$ and for $|k|\lambda\gg1$.
The $-k$ contribution is only of order $10^{-2}$ of the $+k$ contribution [Fig.~\ref{fig:kerr}(d)].
Since $\NlinSC$ depends on the SC only through $\Rz$, it is nearly temperature independent for $T\ll T_c$ (Sec.~\ref{sec:kerr_NL}).
We verify the analytical coefficient by numerically integrating the Landau--Lifshitz (LL) equation on a periodic grid spanning one wavelength within the uniform-thickness approximation and extracting $N_{\rm num}=\Delta\omega/\Delta p_g$ from the mode frequencies obtained by tracking the phase of the fundamental Fourier component at two weak excitation intensities (Appendix~\ref{app:harmonic_balance}).
A direct numerical integration of Eq.~(\ref{eq:LL_classical}) on a periodic grid, with the thin-film dipolar kernel $\mathbf D_0+\mathbf D_{\rm R}$ applied to every Fourier harmonic, reproduces $\Nmag^{(k)}+\NlinSC$ with a relative deviation below $3\times10^{-4}$ at $|k|=1.81\,\mathrm{rad}/\upmu\mathrm m$ for both propagation directions, and below $7\times10^{-4}$ for $0.6\leq|k|\leq5\,\mathrm{rad}/\upmu\mathrm m$ [symbols in Fig.~\ref{fig:linear}(c); Appendix~\ref{app:harmonic_balance}].

\section{Nonlinear Meissner screening}
\label{sec:nonlinear_meissner}

\begin{figure}[!t]
\centering
\includegraphics[width=\columnwidth]{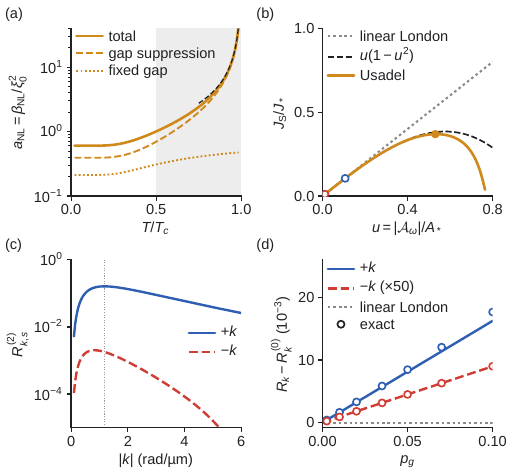}
\caption{Nonlinear Meissner response of dirty Nb.
(a)~Nonlinear coefficient $a_{\rm NL}=\beta_{\rm NL}/\xi_0^2$ from the equilibrium Usadel theory (Appendix~\ref{app:usadel_coeff}), and its contributions at fixed gap and from the self-consistent gap suppression, which equal $2/(3\pi)$ and $\pi/8$ at $T=0$.
Black dashed line: Ginzburg--Landau asymptote $\propto(1-T/T_c)^{-1}$.
The shaded region marks temperatures where applying the equilibrium coefficient to a microwave drive requires a dynamical treatment.
(b)~Supercurrent density versus $u=|\mathcal A_\omega|/A_*=|\mathbf Q|\sqrt{\beta_{\rm NL}}$ from the full Usadel theory at $T=0.1\,T_c$ (solid), the cubic expansion of Eq.~(\ref{eq:microscopic_current_Q}) (dashed), and the linear London relation (dotted).
The filled circle marks the depairing current; open circles mark the interfacial superflow of the $+k$ (blue) and $-k$ (red) modes at $p_g=0.01$.
(c)~Nonlinear reflection slope $\Rt$, Eq.~(\ref{eq:R2_local_main}), for $\pm k$; the vertical line marks $|k|=1.18\,\mathrm{rad}/\upmu\mathrm m$, where $R^{(2)}_{k,+}$ is maximal.
(d)~Amplitude-dependent part of the reflection coefficient at $|k|=1.18\,\mathrm{rad}/\upmu\mathrm m$: first-order result $\Rt p_g$ (lines) and numerical solution of the cubic, fundamental-harmonic London equation (symbols); the $-k$ values are multiplied by 50.
The dotted line is the linear London limit, $R_k=\Rz$.
Panels (b)--(d) use $T=0.1\,T_c$.}
\label{fig:nlme}
\end{figure}

\subsection{Nonlinear London equation}
\label{sec:nonlinear_London}

In a local, vortex-free Meissner state, the supercurrent density $\JS$ of a conventional $s$-wave SC is a function of the gauge-invariant superfluid wave vector
\begin{equation}
 \mathbf Q=\nabla\varphi+\frac{2\pi}{\Phi_0}\mathbf A,
 \label{eq:gauge_invariant_Q}
\end{equation}
where $\varphi$ is the phase of the order parameter, $\Phi_0=h/(2e)$ is the superconducting flux quantum with $h$ the Planck constant and $e$ the elementary charge, and $\hbar\mathbf Q$, with $\hbar=h/2\pi$, is the Cooper-pair momentum.
Locality requires the coherence length $\xi_0$ and the electron mean free path $\ell$ to be short compared with $\lambda$ (Pippard criterion)~\cite{Pippard1953,Tinkham1996}.
Rotational and time-reversal symmetry then imply $\JS\parallel\mathbf Q$ with a magnitude that depends only on $Q^2$, so that for weak superflow
\begin{equation}
 \JS=-\frac{\Phi_0}{2\pi\mu_0\lambda^2}\Big[1-\beta_{\rm NL}(T)\,Q^2\Big]\mathbf Q+O(Q^5).
 \label{eq:microscopic_current_Q}
\end{equation}
The prefactor is the zero-current superfluid stiffness, and the nonlinear coefficient $\beta_{\rm NL}$, with dimension of area, gives the fractional reduction of the superfluid density $n_{\rm s}$ by the flow, $\delta n_{\rm s}/n_{\rm s}=-\beta_{\rm NL}Q^2$.
This is the nonlinear Meissner effect~\cite{Yip1992PRL,Xu1995PRB,Dahm1997JAP,Groll2010NLME,Sauls2022PTEP}.
In circuit language, Eq.~(\ref{eq:microscopic_current_Q}) is a current-dependent kinetic inductance, the origin of the Kerr nonlinearity of superconducting resonators~\cite{Zmuidzinas2012,Eom2012,Swenson2013,Maleeva2018,Shu2021NbTiN}.

The size of $\beta_{\rm NL}$ depends on disorder and on the pairing state.
In a clean $s$-wave SC at $T\rightarrow0$, Galilean invariance leaves $n_{\rm s}$ unchanged up to the Landau critical velocity~\cite{Bardeen1962}, and $\beta_{\rm NL}$ arises only from Doppler-shifted thermal quasiparticles.
It is therefore exponentially small in $\Delta/k_{\rm B}T$, where $\Delta$ is the gap and $k_{\rm B}$ the Boltzmann constant~\cite{Yip1992PRL,Xu1995PRB,Dahm1997JAP}; gap nodes make the correction nonanalytic, $\delta n_{\rm s}/n_{\rm s}\propto|Q|$~\cite{Yip1992PRL,Xu1995PRB}.
In a dirty $s$-wave SC, impurity scattering breaks Galilean invariance, and the superflow acts as a pair-breaking perturbation with depairing energy $\Gamma=\hbar\Dif Q^2/2$, where $\Dif$ is the normal-state diffusion constant~\cite{deGennes1966,Maki1969,Anthore2003,ClemKogan2012,DirtyFlow2020}.
This gives an analytic $Q^2$ correction that remains finite at $T\rightarrow0$.
We consider disordered Nb films ($\ell\lesssim\xi_0$) in this regime and write $\beta_{\rm NL}(T)=a_{\rm NL}(T)\,\xi_0^2$, where $\xi_0=(\hbar\Dif/\Delta_0)^{1/2}$ is the dirty-limit coherence length, $\Delta_0$ is the zero-current gap at $T=0$, and $a_{\rm NL}$ is a dimensionless coefficient.
Expanding the equilibrium Usadel equations~\cite{Usadel1970,Belzig1999} to first order in $\Gamma$ (Appendix~\ref{app:usadel_coeff}) gives
\begin{equation}
 a_{\rm NL}(0)=\frac{2}{3\pi}+\frac{\pi}{8}\simeq0.605.
 \label{eq:aNL0_split}
\end{equation}
The first term describes the loss of condensate spectral weight at a fixed gap; the second stems from the self-consistent gap reduction $\Delta\simeq\Delta_0-\pi\Gamma/4$~\cite{Maki1969,Anthore2003}.
At finite temperature, $a_{\rm NL}=1.005$, $3.109$, and $13.52$ at $T/T_c=0.5$, $0.8$, and $0.95$, approaching the Ginzburg--Landau scaling $\beta_{\rm NL}\propto(1-T/T_c)^{-1}$ [Fig.~\ref{fig:nlme}(a)].

Equation~(\ref{eq:microscopic_current_Q}) is an equilibrium relation. Its use for a microwave drive assumes a quasistatic, predominantly reactive response.
At $T\ll T_c$ and $\hbar\omega\ll2\Delta$, the gap scale supports this approximation and the thermal quasiparticle population is small.
At higher temperatures, the distribution function and the zero- and second-harmonic gap responses should also be considered~\cite{ClemKogan2012,Derendorf2024AC}.
The characteristic electron--phonon time $\tau_0\simeq0.15\,\mathrm{ns}$ quoted for Nb~\cite{Kaplan1976} gives $\omega\tau_0\approx10$ at $11\,\mathrm{GHz}$; it is a warning against assuming instantaneous equilibration, rather than a sample-specific relaxation time.
Nonequilibrium effects modify both the reactive and dissipative response and can include microwave stimulation of superconductivity~\cite{Eliashberg1970,Gurevich2014PRL,deVisser2014PRL,Semenov2020PRApplied}.
We therefore show the high-temperature results as equilibrium estimates, without interpreting them as a rigorous upper bound or the crossover temperature as a universal prediction.

Equation~(\ref{eq:microscopic_current_Q}) defines the vector-potential, current-density, and surface-field scales
\begin{equation}
 A_*=\frac{\Phi_0}{2\pi\sqrt{\beta_{\rm NL}}},\qquad
 J_*=\frac{A_*}{\mu_0\lambda^2},\qquad
 B_*=\frac{A_*}{\lambda},
 \label{eq:Astar_Jstar_Bstar}
\end{equation}
at which the cubic term would cancel the linear one.
They are expansion scales rather than depairing thresholds: in the Ginzburg--Landau regime, the depairing current density is $J_{\rm d}=2J_*/(3\sqrt3)$, and $B_*=\sqrt2B_{\rm c}$ with $B_{\rm c}$ the thermodynamic critical field [Fig.~\ref{fig:nlme}(b)].
For Nb we take $T_c=9.2\,\mathrm K$, $\lambda(0)=226\,\mathrm{nm}$, and $\xi_0=9\,\mathrm{nm}$.
Through the dirty-limit relations $\lambda^{-2}(0)=\pi\mu_0\Delta_0/(\hbar\rho_n)$ and $\xi_0^2=\hbar\Dif/\Delta_0$, with $\Delta_0=1.764\,k_{\rm B}T_c$, these values correspond to a normal-state resistivity $\rho_n\simeq43\,\mu\Omega\,\mathrm{cm}$ and $\Dif\simeq1.7\,\mathrm{cm^2/s}$, an illustrative dirty-film parameter set~\cite{Tinkham1996}.
We obtain $A_*(0)=4.70\times10^{-8}\,\mathrm{T\,m}$, $J_*(0)=7.33\times10^{11}\,\mathrm{A/m^2}$, and $B_*(0)=0.208\,\mathrm T$, close to $B_{\rm c}\simeq0.2\,\mathrm T$ of Nb.
The temperature dependence of $\lambda$ follows from the same theory (Appendix~\ref{app:usadel_coeff}).

For the stray fields of spin waves in the Damon--Eshbach configuration, $\mathbf A=A_z\hat{\mathbf z}$ satisfies $\nabla\cdot\mathbf A=0$ and $\mathbf A\cdot\hat{\mathbf x}=0$ at both SC surfaces, the conditions of the London gauge.
A uniform phase is then consistent; no charge imbalance arises, and $\mathbf Q=(2\pi/\Phi_0)\mathbf A$.
With $A_z=\operatorname{Re}[\mathcal A_\omega(x)\e^{\ii(ky-\omega t)}]$, where $\mathcal A_\omega$ is the complex amplitude at the fundamental frequency, Amp\`ere's law $\nabla^2\mathbf A=-\mu_0\JS$ gives
\begin{equation}
 \big[\partial_x^2-(q_k^{(0)})^2\big]\mathcal A_\omega(x)
 =-\frac{3|\mathcal A_\omega(x)|^2}{4\lambda^2A_*^2}\,\mathcal A_\omega(x).
 \label{eq:nonlinear_London_fundamental}
\end{equation}
The factor $3/4$ is the fundamental Fourier weight of $\cos^3$; the third harmonic at $(3k,3\omega)$ is off resonance with the magnon branch and enters only at higher order~\cite{Derendorf2024AC}.
Because the right-hand side contains the \emph{local} amplitude $\mathcal A_\omega(x)$, nonlinear screening cannot be described by a single amplitude-dependent penetration depth.

The static in-plane field induces an equilibrium screening current along $\hat{\mathbf y}$, orthogonal to the ac current along $\hat{\mathbf z}$.
At fixed dc flow, the cubic expansion first renormalizes the ac stiffness by $\eta_{\rm dc}=\beta_{\rm NL}Q_{\rm dc}^2$, with $Q_{\rm dc}$ the static superfluid wave vector.
Using the linear dc profile gives
\begin{equation}
\eta_{\rm dc}^{\max}=\left[\frac{B_\parallel}{B_*}\tanh\frac{\ts}{2\lambda}\right]^2.
\label{eq:eta_dc}
\end{equation}
For $\ts=65\,\mathrm{nm}$ and $B_\parallel=0.30\,\mathrm T$, this is $0.042$ at $0.1T_c$, which we neglect in the low-temperature example, but rises to $0.071$, $0.220$, and $0.959$ at $T/T_c=0.5$, $0.8$, and $0.95$.
The last values diagnose the breakdown of an expansion about the unbiased condensate, rather than quantitatively describing the biased state.
The parallel lower-critical-field estimate $B_{\rm c1\parallel}\simeq[2\Phi_0/(\pi\ts^2)]\ln(\ts/1.07\xi_0)\approx0.6\,\mathrm T$~\cite{Stejic1994} is compatible with the low-temperature bias for the $65\,\mathrm{nm}$ film.
We additionally assume that residual perpendicular fields and the rf drive do not nucleate vortices.
This is a key advantage of the in-plane geometry over tilted fields, whose perpendicular component generally induces Abrikosov vortices~\cite{Dobrovolskiy2019NP,Dobrovolskiy2025NN}.

\subsection{Interfacial field and nonlinear reflection}
\label{sec:chiral_reflection}

The nonlinear response depends on the field that actually reaches the SC.
The amplitude $\mathcal A_{\rm inc}$ of the vector potential incident on the SC obeys (Appendix~\ref{app:fig1_profile})
\begin{equation}
 \frac{|\mathcal A_{\rm inc}|^2}{p_g}
 =\left(\frac{\BM}{|k|}\right)^2\big(1-\e^{-|k|\tf}\big)^2
 \frac{(1+s\,a_k)^2}{1+a_k^2},
 \label{eq:Ainc_chiral}
\end{equation}
which contains the same directional factor $(1+s\,a_k)^2$ as the linear Meissner shift.
Including the reflected field, the amplitude at the SC surface is $\AS=(1+\Rz)\mathcal A_{\rm inc}$.
We define the interfacial superflow intensity
\begin{equation}
 \begin{aligned}
 \mathcal G_k^{(s)}&\equiv\frac{|\AS|^2}{A_*^2p_g}\\
 &=\frac{|1+\Rz|^2}{1+a_k^2}
 \left[\frac{\BM(1-\e^{-|k|\tf})}{|k|A_*}\right]^2(1+s\,a_k)^2,
 \end{aligned}
 \label{eq:G_definition}
\end{equation}
whose product with $p_g$ is the expansion parameter,
\begin{equation}
 \epsilon_{\rm ac}\equiv\mathcal G_k^{(s)}p_g
=\left(\frac{J_{\rm S}^{(0)}(0)}{J_*}\right)^2
 =\left(\frac{\mu_0h_\parallel(0)}{\bar\Lambda_kB_*}\right)^2.
 \label{eq:epsilon_ac_meaning}
\end{equation}
Here $J_{\rm S}^{(0)}(0)$ and $\mu_0h_\parallel(0)$ are the linear-response amplitudes of the ac supercurrent density and tangential spin-wave field at the interface, and $\bar\Lambda_k=\lambda\Lambda_k^{(0)}\simeq0.5$--$0.7$ for $|k|=1$--$2\,\mathrm{rad}/\upmu\mathrm m$.
The strength of the nonlinear Meissner correction is thus controlled by the interfacial ac field relative to $\bar\Lambda_k B_*$.

We solve Eq.~(\ref{eq:nonlinear_London_fundamental}) perturbatively without replacing the SC field by its surface value.
With the dimensionless coordinate $\zeta=x/\lambda$, wave vector $\bar k=|k|\lambda$, and thickness $d=\ts/\lambda$, let $\psi(\zeta)=\mathcal A_\omega^{(0)}(\zeta)/\AS$ be the linear SC profile normalized at the interface, with $\psi(0)=1$ and $\psi'(d)+\bar k\psi(d)=0$ (Appendix~\ref{app:local_london}).
Multiplying the cubic London equation by $\psi$, integrating across the film, and using the boundary conditions, the dimensionless admittance $\bar\Lambda=-\partial_\zeta\mathcal A_\omega/\mathcal A_\omega|_{\zeta=0}$ acquires the correction
\begin{equation}
 \delta\bar\Lambda_{k,s}=-\frac34\,\epsilon_{\rm ac}I_4,\qquad
 I_j=\int_0^d\psi^j(\zeta)\,\dd\zeta,
 \label{eq:local_Lambda_correction}
\end{equation}
where $I_j$ are overlap integrals of the screening profile. The fourth-order overlap contains three powers from the local cubic current and one from its projection onto the linear field profile.
This is the finite-$k$, finite-thickness counterpart of the nonlinear surface reactance of superconducting films~\cite{Gittleman1965,Dahm1997JAP,Hein1999,Gurevich2014PRL}.
Since $R_k=(\bar k-\bar\Lambda)/(\bar k+\bar\Lambda)$,
\begin{equation}
 R_k(p_g,s)=\Rz+\Rt p_g,\qquad
 \Rt=\frac{3\bar k\,\mathcal G_k^{(s)}I_4}{2(\bar k+\bar\Lambda_k)^2}\geq0,
 \label{eq:R2_local_main}
\end{equation}
up to $O(p_g^2)$, where the superscript $(2)$ indicates second order in the spin-wave amplitude.
Increasing the intensity moves the negative reflection coefficient toward zero, i.e., weakens the screening.
A cruder surface-field approximation, which assigns a uniform amplitude-dependent penetration depth, replaces $I_4$ by $I_2$ and thus overestimates $\Rt$ by $I_2/I_4$, a factor that approaches 2 for a semi-infinite SC.

Figure~\ref{fig:nlme}(c) shows $\Rt$ for spin waves propagating in both directions.
It is maximal for $+k$ at $|k|=1.18\,\mathrm{rad}/\upmu\mathrm m$.
There, the vacuum decay length is $|k|^{-1}\simeq850\,\mathrm{nm}$, the SC penetration length of this Fourier component is $1/q_k^{(0)}\simeq218\,\mathrm{nm}$, and we find $\Rz=-0.329$, $R_{k,+}^{(2)}=0.162$, $R_{k,-}^{(2)}=1.8\times10^{-3}$, and $I_2/I_4=1.13$.
The expansion parameter remains small, $\max_k\epsilon_{\rm ac}=0.013$ at $p_g=0.01$.
The numerical solution of the cubic, fundamental-harmonic Eq.~(\ref{eq:nonlinear_London_fundamental}) deviates from the first-order result by $0.8\%$ at $p_g=0.01$ and by $9\%$ at $p_g=0.1$ [Fig.~\ref{fig:nlme}(d)].

\subsection{Kerr contribution of nonlinear screening}
\label{sec:kerr_NL}

\begin{figure}[!t]
\centering
\includegraphics[width=\columnwidth]{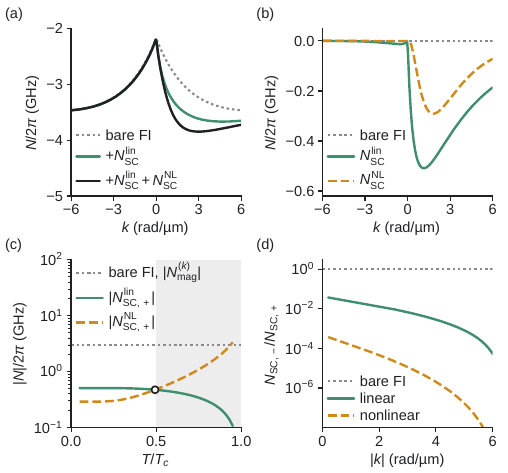}
\caption{Chiral Kerr coefficient of the YIG$|$Nb bilayer (parameters as in Fig.~\ref{fig:linear}).
(a)~Nonlinear frequency-shift coefficient of the bare FI, $\Nmag^{(k)}$ (dotted), of the FI$|$SC bilayer with linear screening only, $\Nmag^{(k)}+\NlinSC$ (green), and with both superconducting channels, $\Ntot$ (black), versus signed $k$ at $T=0.1\,T_c$.
(b)~Linear-screening ($\NlinSC$, solid) and nonlinear Meissner ($\NnlSC$, dashed) contributions at $T=0.1\,T_c$.
(c)~Maxima over $|k|$ of the $+k$ contributions versus temperature, compared with $|\Nmag^{(k)}|$ of the bare FI at $|k|=1.81\,\mathrm{rad}/\upmu\mathrm m$ (dotted).
The open circle marks the crossover at $T\simeq0.49\,T_c$.
The shaded region shows an equilibrium extrapolation: finite-frequency relaxation and dc pair breaking are omitted (Sec.~\ref{sec:nonlinear_London}). The crossover compares separately maximized contributions, not a fixed-$k$ temperature trace.
(d)~Chirality ratios of the two contributions at $T=0.1\,T_c$, Eqs.~(\ref{eq:Nlin_ratio}) and (\ref{eq:NSC_fourth_power_ratio}); the bare FI is reciprocal.}
\label{fig:kerr}
\end{figure}

The amplitude dependence of $R_k$ modifies the frequency through the linear dispersion, Eq.~(\ref{eq:omega_linear}), giving the nonlinear Meissner contribution
\begin{equation}
 \NnlSC(k,s)=\left.\frac{\partial\omega_k^{(0)}}{\partial R_k}\right|_{\Rz}\Rt
 =-\gamma\BM D_k\Rt\,(\nu_k+s),
 \label{eq:Nnl}
\end{equation}
so that the total coefficient [Fig.~\ref{fig:kerr}(a)] is
\begin{equation}
 \Ntot(k,s)=\Nmag^{(k)}+\NlinSC(k,s)+\NnlSC(k,s).
 \label{eq:Ntotal}
\end{equation}
Comparing Eqs.~(\ref{eq:Nlin}) and (\ref{eq:Nnl}), the two superconducting contributions are related by
\begin{equation}
 \frac{\NnlSC}{\NlinSC}=\frac{\Rt}{|\Rz|}\,\frac{\nu_k+s}{\nu_k+s+\chi_k},
 \label{eq:ratio_NL_lin}
\end{equation}
which reduces to $\Rt/|\Rz|$ for the $+k$ branch when $\chi_k\ll\nu_k+1$. The ellipticity correction should be retained for the $-k$ branch.
Both contributions are negative.
Because $\Rt\propto(1+s\,a_k)^2$ and $\nu_k+s\propto(1+s\,a_k)^2$, the nonlinear channel acquires the directional factor twice,
\begin{equation}
 \frac{\NnlSC(-|k|)}{\NnlSC(+|k|)}=\left(\frac{1-a_k}{1+a_k}\right)^4,
 \label{eq:NSC_fourth_power_ratio}
\end{equation}
once from the intensity of the interfacial field that drives the screening currents and once from the sensitivity of the mode frequency to the reflected field.
The linear channel, Eq.~(\ref{eq:Nlin_ratio}), contains only the latter.
The fourth-power law does not depend on the overlap integral $I_4$, which is the same for both directions within the scalar London response.

At $T=0.1\,T_c$, $\NnlSC/2\pi$ peaks at $-0.29~\mathrm{GHz}$ near $|k|=1.8\,\mathrm{rad}/\upmu\mathrm m$ [the dashed line in Fig.~\ref{fig:kerr}(b)], i.e., $10\%$ of $|\Nmag^{(k)}|$.
It peaks at a larger $|k|$ than $\NlinSC$ because the factor $|1+\Rz|$, which controls the field penetrating into the SC, increases with $|k|$.
The contribution of the $-k$ spin waves is suppressed to $10^{-4}$--$10^{-5}$ of that by the $+k$ spin waves [Fig.~\ref{fig:kerr}(d)].
The equilibrium calculation predicts an increasing nonlinear contribution toward $T_c$ [Fig.~\ref{fig:kerr}(c)]. Its maximum over $|k|$ exceeds that of the linear channel near $T\simeq0.49\,T_c$; at $0.8T_c$, the respective maxima are $1.35$ and $0.30\,\mathrm{GHz}$ in magnitude after division by $2\pi$. These high-temperature values omit both finite-frequency relaxation and the growing dc pair breaking quantified by Eq.~(\ref{eq:eta_dc}).
The temperature dependence, directional contrast, and material dependence provide complementary signatures of the two channels: the linear channel involves the SC only through $\lambda$ and $\ts$, whereas the nonlinear channel scales with $\beta_{\rm NL}$.

\section{Directional Kerr compensation}
\label{sec:compensation}

The leading nonlinear frequency shift vanishes when $\Ntot(k,s)=0$.
Since both superconducting contributions are negative, this requires $\Nmag^{(k)}>0$.
For shape-anisotropy-dominated YIG, $\Beff=\BM$ and $\Nmag^{(k)}<0$, so the SC enhances the softening.
According to Sec.~\ref{sec:kerr_bare}, however, $\Nmag^{(k)}$ changes sign when perpendicular anisotropy brings $\Beff$ close to zero.
A garnet film with composition- and strain-tuned perpendicular anisotropy therefore provides a route to compensation while both the magnetization and the bias field remain in plane, subject to the Meissner-state conditions in Sec.~\ref{sec:nonlinear_London}. To isolate the effect of anisotropy tuning, we vary $K_\perp$ while retaining the reference YIG values of $\Ms$ and $A_{\rm ex}$.

\begin{figure}[htp!]
\centering
\includegraphics[width=\columnwidth]{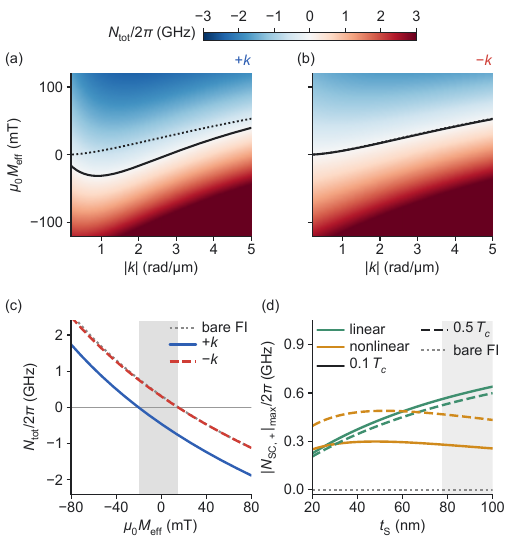}
\caption{Directional Kerr compensation of a garnet film with tunable perpendicular anisotropy, using the reference YIG parameters, $\tf=250\,\mathrm{nm}$, and an in-plane magnetic field $\BH=0.30\,\mathrm T$.
(a,b)~$\Ntot/2\pi$ versus $|k|$ and $\Beff$ for $+k$ and $-k$ at $T=0.1\,T_c$ (color scale clipped at $\pm3\,\mathrm{GHz}$).
Solid lines mark $\Ntot=0$ and dotted lines mark the zero of $\Nmag^{(k)}$ of the bare FI, which in (b) nearly coincides with the solid line.
(c)~Cut at $|k|=1.81\,\mathrm{rad}/\upmu\mathrm m$.
In the shaded window, $-20\,\mathrm{mT}<\Beff<15\,\mathrm{mT}$, counterpropagating spin waves have nonlinear frequency-shift coefficients of opposite sign.
(d)~Maxima over $|k|$ of $|N_{{\rm SC},+}^{\rm lin}|$ (green) and $|N_{{\rm SC},+}^{\rm NL}|$ (amber) versus SC thickness at $T=0.1\,T_c$ (solid) and $0.5\,T_c$ (dashed), for $\Beff=\BM$. The range is restricted to $20$--$100\,\mathrm{nm}$; shading marks $\eta_{\rm dc}^{\max}>0.1$ at $0.5T_c$, a diagnostic of appreciable omitted dc corrections, not a phase boundary. The dashed curves also retain the equilibrium-response assumption.}
\label{fig:comp}
\end{figure}

Figures~\ref{fig:comp}(a,b) map $\Ntot$ in the $(|k|,\Beff)$ plane.
For $-k$, the superconducting contributions are negligible, and the zero contour follows the intrinsic zero at $\Beff\simeq15\,\mathrm{mT}$ for $|k|\simeq1.8\,\mathrm{rad}/\upmu\mathrm m$.
For $+k$, the zero contour is shifted downward by up to $38\,\mathrm{mT}$.
Between the two contours [shaded in Fig.~\ref{fig:comp}(c)], $\Ntot(+|k|)<0<\Ntot(-|k|)$: counterpropagating spin waves acquire nonlinear frequency shifts of opposite sign.
At the window boundaries, the leading shift vanishes for one direction while it remains finite for the other.
The window width is approximately $|\NlinSC+\NnlSC|$ divided by the slope $|\partial\Nmag^{(k)}/\partial\Beff|\simeq0.8\gamma$, and it widens toward $T_c$, subject to the caveats of Sec.~\ref{sec:nonlinear_London}.

The SC thickness controls the two channels differently [Fig.~\ref{fig:comp}(d)].
We restrict this scan to $20$--$100\,\mathrm{nm}$: at $0.1T_c$, $\eta_{\rm dc}^{\max}$ reaches $0.099$ at $100\,\mathrm{nm}$, where the quoted parallel critical-field estimate is already close to $0.30\,\mathrm T$.
At $0.5T_c$, $\eta_{\rm dc}^{\max}=0.1$ is reached near $78\,\mathrm{nm}$; the shaded part of the scan therefore requires a biased-condensate correction.
The magnitude $|\NlinSC|$ grows monotonically with $\ts$, following $|\Rz|$.
The magnitude $|\NnlSC|$ is maximal for $\ts\simeq40$--$60\,\mathrm{nm}$: thinner films screen weakly, whereas in thicker films the penetrating field $|1+\Rz|$ decreases and $I_4/I_2$ approaches $1/2$.
The nonlinear maximum lies within the thin-film part of the scan, where the neglected dc correction is smaller.
Note that at a compensation point only the linear-in-$p_g$ shift vanishes; higher-order shifts, nonlinear damping, and other four-magnon processes generally remain.
At finite $k$, setting $M_{\rm eff}=0$ cancels the uniform anisotropy contribution but leaves dipolar and exchange contributions. The intrinsic Kerr coefficient therefore generally vanishes at a nonzero value of $M_{\rm eff}$.
 
\section{Quantum normalization of the magnon Kerr interaction}
\label{sec:quantum_correspondence}

To relate $\Ntot$ to a magnon Kerr interaction, we quantize the finite-$k$ mode with periodic boundary conditions over a length $L_{\rm r}$ along $y$, chosen so that $kL_{\rm r}/(2\pi)$ is an integer, for a waveguide of width $W$, so that the mode volume is $\Vm=L_{\rm r}W\tf$.
Each magnon of the elliptical mode reduces the longitudinal magnetization by $\gamma\hbar\nu_k/\Vm$, so the classical intensity and the magnon population $n$ are related by
\begin{equation}
 p_g=\frac{\gamma\hbar\nu_k}{\Ms\Vm}\,n.
 \label{eq:power_magnon_mapping}
\end{equation}
We define the Kerr coefficient $\mathcal K$ through the single-mode Hamiltonian
\begin{equation}
 \hat H=\hbar\omega_k^{(0)}\hat n+\hbar\mathcal K\hat n(\hat n-1),\qquad
 \hat n=\hat m^\dagger\hat m,
 \label{eq:Kerr_Hamiltonian}
\end{equation}
where $\hat m$ is the magnon annihilation operator and $\hat n$ the number operator.
Comparing the $n\rightarrow n+1$ transition frequency, $\omega_k^{(0)}+2\mathcal Kn$, with Eq.~(\ref{eq:classical_frequency_expansion}) yields the Kerr coefficient
\begin{equation}
 \mathcal K(k,s)=\frac{\gamma\hbar\nu_k}{2\Ms\Vm}\,\Ntot(k,s).
 \label{eq:K_conversion}
\end{equation}

For the bare magnetic film, the uniform-mode limit makes the dependence on $M_{\mathrm{eff}}$ explicit.
Combining Eqs.~(\ref{eq:NIP_classical}) and (\ref{eq:K_conversion}) gives
\begin{equation}
 \mathcal K_{\mathrm{mag}}^{(0)}
 =-\frac{\gamma^2\hbar\Beff}{4\Ms\Vm}
 \frac{\BH+\Beff/4}{\BH+\Beff}.
 \label{eq:Kmag_uniform}
\end{equation}
This recovers the in-plane result of Ref.~\cite{Petrosyan2026Theory}, expressed using the effective magnetization of Eq.~(\ref{eq:Meff}).
For $\BH>0$ and $\BH+\Beff>0$, its sign is opposite to that of $M_{\mathrm{eff}}$.

At finite $k$, the dipolar and exchange fields also contribute.
Substituting Eqs.~(\ref{eq:selffields_bare}) and (\ref{eq:Nmag_k}) into Eq.~(\ref{eq:K_conversion}) yields
\begin{equation}
 \begin{aligned}
 \mathcal K_{\mathrm{mag}}^{(k)}
 =\frac{\gamma^2\hbar}{16\Ms\Vm}\Big[&-\Beff(1+3a_k^2)\\
 &+3\BM(1-F_k)(a_k^2-a_k^{-2})\\
 &+\Bex(a_k^2+a_k^{-2}-10)\Big].
 \end{aligned}
 \label{eq:Kmag_finite_k}
\end{equation}
The three terms arise from the effective anisotropy, finite-$k$ dipolar fields, and exchange, respectively.
The dependence on $M_{\mathrm{eff}}$ also enters through $a_k=\sqrt{\mathcal H_1/\mathcal H_2}$.
We denote the bare-film value by $\mathcal K_{\mathrm{bare}}(k)\equiv\left.\mathcal K_{\mathrm{mag}}^{(k)}\right|_{\Rz=0}$, with the ellipticity evaluated from the unscreened stiffness fields of Eq.~(\ref{eq:omega_bare}).

For the YIG parameters of Fig.~\ref{fig:linear}, $M_{\mathrm{eff}}=\Ms$, $|k|=1.81\,\mathrm{rad}/\upmu\mathrm m$, and the reference volume $\Vm=25\,\upmu\mathrm m^3$ give $\mathcal K_{\mathrm{bare}}(k)/(2\pi)=-5.76\,\mathrm{mHz}$ for both propagating directions of spin waves.

Adding the SC gives $\mathcal K=\mathcal K_{\mathrm{mag}}^{(k)}+\mathcal K_{\mathrm{SC}}^{\mathrm{lin}}+\mathcal K_{\mathrm{SC}}^{\mathrm{NL}}$, where
\begin{equation}
 \begin{aligned}
 \mathcal K_{\mathrm{SC}}^{\mathrm{lin}}(k,s)
 &=\frac{\gamma^2\hbar\nu_k}{2\Ms\Vm}\,G_k(\nu_k+s+\chi_k),\\
 \mathcal K_{\mathrm{SC}}^{\mathrm{NL}}(k,s)
 &=-\frac{\gamma^2\hbar\nu_k}{2\Ms\Vm}\,\BM D_k\Rt(\nu_k+s),
 \end{aligned}
 \label{eq:K_SC_channels}
\end{equation}
where $a_k$, $\nu_k$, and $\chi_k$ are evaluated using the screened stiffness fields of Eq.~(\ref{eq:omega_linear}), including in $\mathcal K_{\mathrm{mag}}^{(k)}$.
These expressions apply to either sign of $M_{\mathrm{eff}}$ within the present model, provided the in-plane mode remains stable, with $\mathcal H_{1,2}>0$ and $\omega_k^{(0)}(s)>0$.
The bare-film limit follows by setting $\Rz=\Rt=0$ and using the corresponding unscreened ellipticity.

For the same YIG mode covered by $65\,\mathrm{nm}$ Nb at $T=0.1\,T_c$, the linear and nonlinear Meissner screening contributes to $\mathcal K/(2\pi)$ are $-0.92$ and $-0.56\,\mathrm{mHz}$ for $+k$, respecitively.
For $-k$, they become negligible with $-0.014$ and $-3.2\times10^{-5}\,\mathrm{mHz}$.
Including the change in the magnetic term due to the screened ellipticity, the totals are $-7.26$ and $-5.79\,\mathrm{mHz}$, respectively.
Thus, the $-k$ total remains close to the bare-film value because its superconducting corrections are small.

At the anisotropy-compensation point $M_{\mathrm{eff}}=0$, retaining the other reference parameters gives a bare-film coefficient $\mathcal K_{\mathrm{bare}}(k)/(2\pi)=+0.622\,\mathrm{mHz}$. With Nb, the total $\mathcal K/(2\pi)$ becomes $-0.874\,\mathrm{mHz}$ for $+k$ and $+0.593\,\mathrm{mHz}$ for $-k$, illustrating the directional sign control in Fig.~\ref{fig:comp}.
All these coefficients scale as $1/\Vm$ and refer to the thickness-uniform mode. For the quoted volume, $p_g=10^{-3}$ corresponds to $n\simeq2.6\times10^8$ at $M_{\rm eff}=\Ms$: a millihertz coefficient per magnon therefore produces a megahertz frequency shift at a large coherent population.
A standing-wave mode also involves the cross-Kerr coupling between its $+k$ and $-k$ components, which requires their four-magnon matrix element~\cite{Krivosik2010,Bruehlmann2026Continuum}.

\section{Discussion and conclusion}
\label{sec:conclusion}

The predicted shifts can be tested by power-dependent resonance or propagation measurements.
For the thickness-uniform YIG example at $|k|=1.81\,\mathrm{rad}/\upmu\mathrm m$, the two explicit superconducting channels give a $+k$ softening of approximately $0.8$--$8\,\mathrm{MHz}$ at $p_g=10^{-3}$--$10^{-2}$, comparable to low-temperature YIG/GGG linewidths~\cite{Kosen2019}.
The accessible intensity should remain below the onset of additional spin-wave instabilities~\cite{AndersonSuhl1955,Suhl1957,GurevichMelkov1996}.
Thin-film Kerr measurements using microwave resonators~\cite{Petrosyan2026Exp} provide one route, provided the measured mode profile and normalization are included.

Propagating spin-wave spectroscopy~\cite{Vlaminck2008} measures the phase at a fixed drive frequency.
With the phase convention $\exp[\ii(ky-\omega t)]$ and distance $l$ measured along the propagation direction, the local wave-number change is $\delta|k|=-\Ntot(s)p_g/v_{g,s}$, where $v_{g,s}=\partial\omega_k^{(0)}(s)/\partial|k|$.
The accumulated phase is therefore
\begin{equation}
 \Delta\phi_s=-\frac{\Ntot(s)}{v_{g,s}}\int_0^{L_{\rm p}}p_g(l)\,\dd l.
 \label{eq:propagation_phase}
\end{equation}
For constant $p_g=10^{-3}$ and $L_{\rm p}=20\,\upmu\mathrm m$, the explicit superconducting channels contribute $+0.34\,\mathrm{rad}$ for $+k$ and $+1.25\times10^{-3}\,\mathrm{rad}$ for $-k$ in the uniform model.
Their group velocities differ, $v_{g,+}=0.286$ and $v_{g,-}=0.725\,\mathrm{km/s}$, so a raw phase difference does not cancel the common magnetic contribution.
Each phase should first be converted to $\Ntot(s)$ using its measured group velocity and calibrated integrated intensity; the two slopes can then be compared at equal $|k|$.

A bare-film comparison should also include the change in ellipticity. The complete superconductivity-induced change is
\begin{equation}
 \begin{aligned}
 \Delta N_{\rm SC}={}&\Nmag^{(k)}(a_k^{\rm SC})-\Nmag^{(k)}(a_k^{\rm bare})\\
 &+\NlinSC+\NnlSC,
 \end{aligned}
 \label{eq:measured_SC_change}
\end{equation}
where the superscripts identify the screened and bare evaluations of $a_k$.
At the same wave vector, the first difference is $-2\pi\times7.00\,\mathrm{MHz}$, comparable to the $-k$ linear-screening term.
Within the uniform model this magnetic term is reciprocal and cancels in the difference of the reconstructed slopes.
After calibrating the linear dispersion, the distinct temperature and thickness dependencies can help isolate nonlinear screening. The predicted $0.49T_c$ crossover is an equilibrium benchmark, rather than a model-independent experimental identifier.

Our low-temperature calculation assumes weak ac and dc superflow, a local response with $\xi_0,\ell\ll\lambda$, and predominantly reactive conductivity, $\sigma_1/\sigma_2\ll1$~\cite{Mattis1958}.
For the $65\,\mathrm{nm}$ Nb example at $0.1T_c$, $\epsilon_{\rm ac}\leq0.013$ at $p_g=0.01$ and $\eta_{\rm dc}^{\max}=0.042$.
These bounds do not apply to the entire temperature or thickness scan: near $T_c$, finite-frequency relaxation, dc pair breaking, and nonlinear damping require a treatment about the biased nonequilibrium state~\cite{ClemKogan2012,KoganProzorov2020AC,Gurevich2014PRL,Derendorf2024AC}.
Patterning a finite superconducting gate additionally introduces lateral boundaries and scattering, which are absent from the translationally invariant bilayer considered here.

In conclusion, Meissner screening modifies the magnon Kerr interaction through two chiral mechanisms.
Linear screening combines a linear reflected field with the fixed-length constraint on the magnetization; nonlinear screening makes the reflected field itself amplitude dependent.
Their different directional factors and parameter dependencies provide complementary means of separating them.
For the illustrative low-temperature bilayer, anisotropy tuning produces opposite Kerr signs for counterpropagating modes within the uniform-thickness approximation.
These results establish a mechanism for directional control of magnon interactions within the stated Meissner regime.

\begin{acknowledgments}
H.W. acknowledges the support of the Swiss National Science Foundation (Grant No.~10004526).
T.Y. is financially supported by the National Natural Science Foundation of China under Grant No.~12374109 and the National Key Research and Development Program of China under Grant No.~2023YFA1406600.
\end{acknowledgments}

%\clearpage
\appendix

\section{Interfacial field and magnetic-field profiles}
\label{app:fig1_profile}

The stray field of the FI is generated by magnetic surface charges with complex amplitudes $\pm\mathcal M_x$ at $x=0$ and $x=-\tf$ and by a volume charge with amplitude $-\ii k\,\mathcal M_y$.
With the vector-potential convention of Eq.~(\ref{eq:app_B_components}), the incident amplitude is
\begin{equation}
 \mathcal A_{\rm inc}=-\frac{\mu_0}{2|k|}\big(1-\e^{-|k|\tf}\big)
 \big(\mathcal M_y+\ii s\,\mathcal M_x\big).
 \label{eq:app_Ainc}
\end{equation}
Here $\mathcal M_x=-\mathcal M_2$ and $\mathcal M_y=\mathcal M_1$ are the Cartesian magnetization amplitudes defined in Sec.~\ref{sec:linear_bare}.
The identity $\mathcal M_y+\ii s\,\mathcal M_x=\ii s(\mathcal M_x-\ii s\,\mathcal M_y)$ relates this expression to the combination that couples to $\mathbf D_{\rm R}$ in Eq.~(\ref{eq:DR}).
Below the film, the corresponding combination is $\mathcal M_x+\ii s\,\mathcal M_y$.
For the positive-frequency mode, $\mathcal M_2=\ii a_k\mathcal M_1$, so that $|\mathcal M_x\mp\ii s\,\mathcal M_y|=|\mathcal M_1|\,|1\pm s\,a_k|$.
Together with Eq.~(\ref{eq:pg}), this yields Eq.~(\ref{eq:Ainc_chiral}).

Figure~\ref{fig:principle}(c,d) shows the resulting near field.
The thickness-uniform magnetization is equivalent to bound volume and surface currents with complex amplitudes $J_z=-\ii k\,\mathcal M_x$, $K_z(0)=-\mathcal M_y$, and $K_z(-\tf)=+\mathcal M_y$.
In free space, their vector-potential amplitude is
\begin{equation}
 \begin{aligned}
 \mathcal A_0(x)=\frac{\mu_0}{2|k|}\Big[&\int_{-\tf}^{0}J_z(x')\,\e^{-|k||x-x'|}\dd x'\\
 &+K_z(0)\e^{-|k||x|}\\
 &+K_z(-\tf)\e^{-|k||x+\tf|}\Big].
 \end{aligned}
 \label{eq:app_A0_currents}
\end{equation}
At the upper interface, $\mathcal A_0(0)=\mathcal A_{\rm inc}$ reproduces Eq.~(\ref{eq:app_Ainc}), and the incident field above the FI is $\mathcal A_{\rm inc}\e^{-|k|x}$.
In the bilayer, the vector potential is $\mathcal A_0(x)+\Rz\mathcal A_{\rm inc}\e^{|k|x}$ for $x<0$, $(1+\Rz)\mathcal A_{\rm inc}$ times the profile of Eq.~(\ref{eq:app_SC_solution}) normalized at $x=0$ inside the SC, and decays as $\exp[-|k|(x-\ts)]$ above the SC.
Because $\mathbf B=\nabla\times(A_z\hat{\mathbf z})$, the field lines in Fig.~\ref{fig:principle}(c) are the contours of $A_z(x,y)=\operatorname{Re}[\mathcal A(x)\e^{\ii ky}]$, and the screening current in the SC is $J_{{\rm S},z}=-A_z/(\mu_0\lambda^2)$.
Below the FI, the complex amplitude is proportional to $(1-s\,a_k)-\Rz(1+s\,a_k)\e^{-|k|\tf}$: for $+k$, the Meissner image of the strong upper-surface field also emerges below the film, whereas for $-k$ the SC-side amplitude, and with it the screening current, is reduced by $|1-a_k|/(1+a_k)\approx0.1$ relative to $+k$.

\section{Derivation of the Meissner reflection coefficient}
\label{app:linear_reflection}

We derive the linear magnetic-field reflection coefficient of a superconducting film with finite thickness $\ts$.
The FI--SC interface is located at $x=0$, the SC occupies $0<x<\ts$, and the region $x>\ts$ is vacuum.
We consider a Fourier component with in-plane wave vector $k$ along $y$ and suppress its common factor $\exp[\ii(ky-\omega t)]$.
With $\mathbf A=\mathcal A(x)\hat{\mathbf z}$, the magnetic induction has the components
\begin{equation}
 B_x=\ii k\mathcal A,\qquad B_y=-\partial_x\mathcal A.
 \label{eq:app_B_components}
\end{equation}
In the quasistatic limit, the Maxwell equation $\nabla\times\mathbf B=\mu_0\JS$, together with the linear London relation $\JS=-\mathbf A/(\mu_0\lambda^2)$, gives Eq.~(\ref{eq:qlinear}).

Since there are no magnetic sources for $x>\ts$, the vacuum solution that vanishes at infinity is $\mathcal A(\ts)\exp[-|k|(x-\ts)]$.
Continuity of the vector potential and of the tangential magnetic field at $x=\ts$, where no surface-current sheet is present, requires $\partial_x\mathcal A|_{x=\ts^-}=-|k|\mathcal A(\ts)$.
The solution inside the SC satisfying this boundary condition is
\begin{equation}
 \mathcal A_{\rm SC}(x)=\mathcal A(\ts)\Big[\cosh q_k^{(0)}(\ts-x)
 +\frac{|k|}{q_k^{(0)}}\sinh q_k^{(0)}(\ts-x)\Big],
 \label{eq:app_SC_solution}
\end{equation}
whose logarithmic derivative at the FI--SC interface reproduces Eq.~(\ref{eq:Lambdafinite}) and establishes the effective boundary condition $\partial_x\mathcal A_{\rm SC}(0)=-\Lambda_k^{(0)}\mathcal A_{\rm SC}(0)$.

On the FI side of the interface, the source-free field can be written as
\begin{equation}
 \mathcal A(x)=\mathcal A_{\rm inc}\e^{-|k|x}+\mathcal A_{\rm ref}\e^{|k|x},
 \label{eq:app_incident_reflected}
\end{equation}
where $\mathcal A_{\rm inc}$ is the amplitude of the field incident on the SC and $\mathcal A_{\rm ref}$ that of the field reflected back toward the FI.
Matching $\mathcal A$ and $\partial_x\mathcal A$ to the effective boundary condition gives $\Rz\equiv\mathcal A_{\rm ref}/\mathcal A_{\rm inc}$ in the form of Eq.~(\ref{eq:Rlinear}), and the total vector potential at the SC surface $\mathcal A_{\rm SC}(0)=(1+\Rz)\mathcal A_{\rm inc}$.
Several limiting cases provide consistency checks.
For a semi-infinite SC, $\ts\rightarrow\infty$, one finds $\Lambda_k^{(0)}\rightarrow q_k^{(0)}$ and $\Rz\rightarrow(|k|-q_k^{(0)})/(|k|+q_k^{(0)})$.
In the absence of superconducting screening, $\lambda^{-1}\rightarrow0$, or for $\ts\rightarrow0$, one has $\Lambda_k^{(0)}\rightarrow|k|$ and hence $\Rz\rightarrow0$.

\section{Harmonic balance for the finite-$k$ mode}
\label{app:harmonic_balance}

Here we derive Eqs.~(\ref{eq:Nmag_k}) and (\ref{eq:Nlin}) from Eq.~(\ref{eq:HB_SC}); setting $G_k=0$ recovers the bare-film result of Eq.~(\ref{eq:HB_bare}).
The reduction factors are $c_{x,y}=1-\delta_{x,y}$, with
\begin{equation}
 \begin{aligned}
 \delta_x&=\frac{3\bar m_x^2+\bar m_y^2}{8}=\frac{(3a_k^2+1)\,p_g}{2(1+a_k^2)},\\
 \delta_y&=\frac{\bar m_x^2+3\bar m_y^2}{8}=\frac{(a_k^2+3)\,p_g}{2(1+a_k^2)},
 \end{aligned}
 \label{eq:app_delta}
\end{equation}
where we used $\bar m_x^2=4p_ga_k^2/(1+a_k^2)$ and $\bar m_y^2=4p_g/(1+a_k^2)$, and the second-harmonic amplitude is $\Xi=2p_g(a_k^2-1)/(1+a_k^2)$.
Multiplying the two lines of Eq.~(\ref{eq:HB_SC}) gives
\begin{equation}
 \begin{aligned}
 &\big(\omega+\gamma c_ysG_k\big)\big(\omega+\gamma c_xsG_k\big)\\
 &\quad=\gamma^2\big(\Hone+\delta_yb_1-\Bex\Xi\big)\\
 &\qquad\times\big(\Htwo+\delta_xb_2+\Bex\Xi\big).
 \end{aligned}
 \label{eq:app_HB_product}
\end{equation}
To first order in $p_g$, the left-hand side equals $[\omega+\gamma sG_k(1-p_g)]^2$, since $\delta_x+\delta_y=2p_g$ and the difference $c_x-c_y$ contributes only at $O(p_g^2)$.
On the right-hand side we use $\Hone=a_k^2\Htwo$ and expand the square root.
This yields
\begin{align}
 \frac{\omega}{\gamma}&=\sqrt{\Hone\Htwo}-sG_k\nonumber\\
 &+p_g\bigg[\frac{(a_k^2+3)b_1+a_k^2(3a_k^2+1)b_2}{4a_k(1+a_k^2)}\nonumber\\
 &+sG_k+\Bex\frac{(1-a_k^2)^2}{a_k(1+a_k^2)}\bigg].
 \label{eq:app_HB_result}
\end{align}
Separating $b_j=b_j^{\rm m}+G_k$ and using $[(a_k^2+3)+a_k^2(3a_k^2+1)]/[4a_k(1+a_k^2)]=\nu_k+\chi_k$ gives Eqs.~(\ref{eq:Nmag_k}) and (\ref{eq:Nlin}).

We verified Eq.~(\ref{eq:app_HB_result}) by integrating the full Landau--Lifshitz equation on a periodic grid spanning one wavelength.
The thin-film dipolar kernel $\mathbf D_0+\mathbf D_{\rm R}$ was applied to every Fourier harmonic, exchange was included for all components, and the amplitude dependence of the frequency was extracted from the phase velocity of traveling waves with $p_g=0.002$ and $0.006$.
For the parameters of Fig.~\ref{fig:linear} at $|k|=1.81\,\mathrm{rad}/\upmu\mathrm m$, the numerical and analytical values of $\Nmag^{(k)}+\NlinSC$ agree with a relative deviation below $3\times10^{-4}$ for both propagation directions, and the linear frequencies to within a relative deviation of $10^{-6}$.
Over $0.6\leq|k|\leq5\,\mathrm{rad}/\upmu\mathrm m$ [symbols in Fig.~\ref{fig:linear}(c)], the relative deviation stays below $7\times10^{-4}$, consistent with the $O(p_g)$ curvature of the finite-amplitude frequency.

\section{Equilibrium dirty-limit nonlinear coefficient}
\label{app:usadel_coeff}

The nonlinear current coefficient is obtained from the equilibrium Usadel equations for a spatially uniform, vortex-free dirty $s$-wave superconductor~\cite{Usadel1970,Belzig1999,ClemKogan2012,Sauls2022PTEP}.
We parametrize the normal and anomalous Matsubara Green functions as $g_n=\cos\vartheta_n$ and $f_n=\sin\vartheta_n$, where $\vartheta_n$ is the pairing angle at the fermionic Matsubara energy $\varepsilon_n=(2n+1)\pi k_{\rm B}T$ with integer $n\geq0$.
The Usadel equation then reads
\begin{equation}
 (\varepsilon_n+\Gamma\cos\vartheta_n)\sin\vartheta_n=\Delta\cos\vartheta_n,
 \label{eq:usadel_theta}
\end{equation}
where $\Gamma=\hbar\Dif Q^2/2$ is the superflow pair-breaking energy.
The gap $\Delta=\Delta(Q,T)$ obeys the self-consistency equation
\begin{equation}
 \Delta\ln\frac{T}{T_c}+2\pi k_{\rm B}T\sum_{n\geq0}
 \left(\frac{\Delta}{\varepsilon_n}-\sin\vartheta_n\right)=0,
 \label{eq:usadel_gap}
\end{equation}
and the supercurrent satisfies $J_{\rm S}(Q,T)/Q\propto-T\sum_{n\geq0}\sin^2\vartheta_n(Q,T)$.
The nonlinear coefficient is extracted from the same current--momentum relation as
\begin{equation}
 \beta_{\rm NL}(T)=-\left.\frac{\partial_{Q^2}[J_{\rm S}(Q,T)/Q]}{J_{\rm S}(Q,T)/Q}\right|_{Q=0},
 \label{eq:beta_extract}
\end{equation}
in which the proportionality constant cancels.

To include the suppression of the gap by superflow, let $\Delta_T=\Delta(0,T)$ be the zero-current gap at temperature $T$, $E_n=(\varepsilon_n^2+\Delta_T^2)^{1/2}$, and expand $\Delta(Q,T)=\Delta_T+\Gamma\Delta_1(T)+O(\Gamma^2)$, where $\Delta_1$ is the dimensionless slope of the gap with respect to $\Gamma$.
Expanding Eq.~(\ref{eq:usadel_gap}) gives
\begin{equation}
 \Delta_1(T)=-\frac{\sum_{n}\varepsilon_n^2/E_n^4}{\Delta_T\sum_{n}1/E_n^3},
 \label{eq:usadel_gap_slope}
\end{equation}
and Eq.~(\ref{eq:usadel_theta}) yields
\begin{equation}
 \sin\vartheta_n=\frac{\Delta_T}{E_n}
 +\Gamma\frac{\varepsilon_n^2}{E_n^4}\big(E_n\Delta_1-\Delta_T\big)+O(\Gamma^2).
 \label{eq:usadel_theta_expand}
\end{equation}
Defining the two Matsubara sums
\begin{equation}
 S_0=\sum_{n}\frac{\Delta_T^2}{E_n^2},\qquad
 S_1=2\sum_{n}\left(\frac{\Delta_T\Delta_1\varepsilon_n^2}{E_n^4}
 -\frac{\Delta_T^2\varepsilon_n^2}{E_n^5}\right),
 \label{eq:usadel_sums}
\end{equation}
we obtain $[J_{\rm S}(Q,T)/Q]/\lim_{Q\to0}[J_{\rm S}(Q,T)/Q]=1+\Gamma S_1/S_0+O(\Gamma^2)$.
With $\xi_0^2=\hbar\Dif/\Delta_0$, the nonlinear coefficient becomes
\begin{equation}
 \beta_{\rm NL}(T)=-\frac{\hbar\Dif}{2}\frac{S_1}{S_0},\qquad
 a_{\rm NL}(T)=-\frac{\Delta_0}{2}\frac{S_1}{S_0}.
 \label{eq:usadel_aNL_finite_T}
\end{equation}

At $T=0$, the Matsubara sums become integrals, and Eq.~(\ref{eq:usadel_gap_slope}) gives $\Delta_1(0)=-\pi/4$.
The resulting current correction is $\lim_{T\to0}S_1/S_0=-\big[\pi/4+4/(3\pi)\big]/\Delta_0$, so that $a_{\rm NL}(0)=\pi/8+2/(3\pi)\simeq0.6049$.
The first term arises from the self-consistent reduction of $\Delta$, whereas the second survives if the gap is held fixed.
For weak-coupling BCS theory, $\Delta_0\simeq1.764\,k_{\rm B}T_c$, and numerical evaluation gives $a_{\rm NL}=0.6049$, $1.0054$, $3.1091$, and $13.519$ at $T/T_c=0.1$, $0.5$, $0.8$, and $0.95$, respectively.
The same calculation yields the penetration depth through $\lambda^{-2}(T)/\lambda^{-2}(0)=(\Delta_T/\Delta_0)\tanh[\Delta_T/(2k_{\rm B}T)]$.
These equilibrium coefficients assume a homogeneous, unbiased condensate; finite-frequency order-parameter dynamics, spatially varying dc screening currents, and vortices require separate treatment.

\section{Spatially resolved nonlinear London response}
\label{app:local_london}

Here we derive the spatially resolved result of Eq.~(\ref{eq:local_Lambda_correction}).
With the dimensionless variables $\zeta=x/\lambda$, $\bar k=|k|\lambda$, and $d=\ts/\lambda$ of Sec.~\ref{sec:chiral_reflection}, and with $\bar q=(1+\bar k^2)^{1/2}$ the dimensionless inverse penetration length, the linear SC profile normalized at $\zeta=0$ is
\begin{equation}
 \psi(\zeta)=\frac{\e^{-\bar q\zeta}+\varrho_k\e^{-\bar q(2d-\zeta)}}
 {1+\varrho_k\e^{-2\bar qd}},\qquad
 \varrho_k=\frac{\bar q-\bar k}{\bar q+\bar k},
 \label{eq:app_phi}
\end{equation}
where $\varrho_k$ is the internal reflection amplitude at the outer SC surface.
This profile obeys $\psi(0)=1$ and $\psi'(d)+\bar k\psi(d)=0$.
For a real harmonic fundamental, Eq.~(\ref{eq:nonlinear_London_fundamental}) reads
\begin{equation}
 \psi_{\rm NL}''-\bar q^2\psi_{\rm NL}=-\frac34\epsilon_{\rm ac}\,\psi_{\rm NL}^3,
 \label{eq:app_local_equation}
\end{equation}
where $\psi_{\rm NL}(\zeta)=\mathcal A_\omega(\zeta)/\AS$ is the normalized nonlinear profile and $\epsilon_{\rm ac}$ is given by Eq.~(\ref{eq:epsilon_ac_meaning}).
For a fixed surface amplitude, the correction $\delta\psi=\psi_{\rm NL}-\psi$ obeys $\delta\psi(0)=0$ and the same outer boundary condition.
Multiplying its linearized equation by $\psi$, integrating over the SC thickness, and using these boundary conditions gives
\begin{equation}
 \delta\bar\Lambda=-\delta\psi'(0)=-\frac34\epsilon_{\rm ac}I_4,
 \label{eq:app_dLambda}
\end{equation}
which is Eq.~(\ref{eq:local_Lambda_correction}).
Since $R_k=(\bar k-\bar\Lambda)/(\bar k+\bar\Lambda)$, Eq.~(\ref{eq:R2_local_main}) follows.
The use of the linear surface amplitude in $\epsilon_{\rm ac}$ is sufficient at this order, because its nonlinear correction enters at higher order in $p_g$.

For the uniform surface-field approximation, $\partial_{\bar q}\bar\Lambda=2\bar qI_2$, so that its correction has the same form as Eq.~(\ref{eq:R2_local_main}) with $I_4$ replaced by $I_2$.
For a semi-infinite SC, $\psi(\zeta)=\e^{-\bar q\zeta}$ and $I_4/I_2=1/2$, so the surface-field approximation overestimates the nonlinear reflection even as $p_g\rightarrow0$.
A finite dc screening-current profile alters the linear operator and requires a new perturbative solution around the biased equilibrium.

\ifrevtexavailable\else
\bibliographystyle{unsrt}
\fi
\bibliography{references}

\end{document}